\documentclass[aps,twocolumn,pra,groupedaddress,superscriptaddress,nofootinbib]{revtex4-2}

\usepackage{graphicx,amsmath,amssymb,amsbsy,subfigure,hyperref,bbm,times,txfonts,float}
\usepackage{subfigure,hyperref,bbm,times}
\usepackage[T1]{fontenc}
\usepackage{braket}
\usepackage{epsfig} 
\usepackage{color}
\usepackage{graphicx}% Include figure files
\usepackage{dcolumn}
\usepackage{bm}% bold math

\providecommand{\openone}{\leavevmode\hbox{\small1\kern-3.8pt\normalsize1}}

\usepackage{soul}

\hypersetup{
   colorlinks=true,
   linkcolor=blue,
}
\graphicspath{{figure/}}

\begin{document}

%\preprint{APS/123-QED}

\title{Preserving quantum coherence in thermal noisy systems via qubit frequency modulation}

\author{Mahshid Khazaei Shadfar}
\email{mahshid.shadfar@gmail.com}
\affiliation{Dipartimento di Ingegneria, Universit\`{a} di Palermo, Viale delle Scienze, 90128 Palermo, Italy}
\affiliation{INRS-EMT, 1650 Boulevard Lionel-Boulet, Varennes, Qu\'{e}bec J3X 1S2, Canada}

\author{Farzam Nosrati}
\affiliation{Dipartimento di Ingegneria, Universit\`{a} di Palermo, Viale delle Scienze, 90128 Palermo, Italy}
\affiliation{IMDEA Networks Institute, Madrid, Spain}

\author{Ali Mortezapour}
\affiliation{Department of Physics, University of Guilan, P. O. Box 41335--1914, Rasht, Iran}

\author{Vincenzo Macri}
\affiliation{Dipartimento di Fisica, Università di Pavia, via Bassi 6 , 27100 Pavia l, Italy}

\author{Roberto Morandotti}
\affiliation{INRS-EMT, 1650 Boulevard Lionel-Boulet, Varennes, Qu\'{e}bec J3X 1S2, Canada}

\author{Rosario Lo Franco} 
\email{rosario.lofranco@unipa.it}
\affiliation{Dipartimento di Ingegneria, Universit\`{a} di Palermo, Viale delle Scienze, 90128 Palermo, Italy}

\begin{abstract}

Quantum coherence is a key resource underpinning quantum technologies, yet it is highly susceptible to environmental decoherence, especially in thermal settings. While frequency modulation (FM) has shown promise in preserving coherence at zero temperature, its effectiveness in realistic, noisy thermal environments remains unclear. In this work, we investigate a single frequency-modulated qubit interacting with a thermal phase-covariant reservoir composed of dissipative and dephasing channels. We demonstrate that FM significantly preserves coherence in the presence of thermal dissipation while being ineffective under thermal pure-dephasing noise due to commutation between system and interaction Hamiltonians. When both noise channels are present, FM offers protection only for weak dephasing coupling. Our findings clarify the limitations and potential of FM-based coherence protection under thermal noise, supplying practical insights into designing robust quantum systems for quantum applications.

\end{abstract}

%\date{\today }

\maketitle

\section{Introduction}

Quantum coherence, a direct consequence of the superposition principle, is a fundamental aspect of quantum mechanics that sets it apart from classical mechanics \cite{leggett1980suppl, yao2015quantum, streltsov2015measuring, hu2016quantum, radhakrishnan2016distribution, ma2016converting, malvezzi2016quantum, li2016quantum, chen2016coherence, hu2017relative, hu2018quantum}. Beyond its foundational significance, quantum coherence is also a crucial resource for various quantum technologies and protocols \cite{chitambar2016comparison, napoli2016robustness, rana2016trace, yu2016alternative, streltsov2017colloquium, baumgratz2014quantifying, winter2016operational, chitambar2016relating}. These include quantum metrology \cite{giovannetti2004quantum, demkowicz2014using,SunPNAS}, nonclassicality measures \cite{asboth2005computable, streltsov2015measuring}, quantum dense coding \cite{bennett1992communication}, quantum error correction (QEC) \cite{laflamme1996perfect, plenio1997quantum}, quantum key distribution (QKD) \cite{ekert1991quantum}, quantum computing \cite{unruh1995maintaining, duan1997preserving}, and quantum thermalization processes \cite{wang2001entanglement, arnesen2001natural}. By leveraging quantum coherence, these protocols can enhance precision \cite{giovannetti2011advances, joo2011quantum, liu2013phase, chaves2013noisy, zhang2013quantum, demkowicz2014using, alipour2014quantum, lu2015robust, correa2015individual, baumgratz2016quantum, liu2017quantum} and computational efficiency \cite{liu2023application, shahandeh2019quantum}.

However, in realistic settings, any quantum system inevitably interacts with its surrounding environment, leading to decoherence, that is the degradation of coherence due to the loss of phase information between components of a quantum superposition \cite{breuer2002theory, zurek1991decoherence, zurek1994decoherence}. Numerous strategies have been proposed to mitigate decoherence and protect quantum coherence \cite{lofranco2013dynamics, mortezapour2018coherence, scala2008population, duan1997preserving, viola1998dynamical, branderhorst2008coherent, franco2013spin, tan2010non, scala2011robust, xue2012decoherence, d2014recovering, man2012enhancing}. Among these, frequency modulation (FM) of the qubit’s transition frequency has emerged as a promising signal-processing technique for decoherence suppression. In this approach, the qubit’s transition frequency $\omega_0$ is modulated sinusoidally by an external off-resonant field \cite{noel1998frequency, zhang2003frequency, silveri2017quantum}, which can effectively decouple the system from its reservoir.

Recent studies have shown that frequency-modulated qubits exhibit a rich variety of quantum phenomena, including the dynamic Stark effect \cite{alsing1992dynamic}, Landau-Zener-Stückelberg interference \cite{shevchenko2010landau}, enhanced non-Markovianity \cite{poggi2017driving}, topological transitions \cite{martin2017topological}, and population trapping \cite{gray1978coherent}, highlighting the increasing relevance of this technique.

Most quantum protection schemes have been investigated under the assumption of zero temperature, where thermal noise is absent \cite{mortezapour2018coherence, scala2008population, carrion2024decoherence, huang2017optimal}. However, in practical scenarios, quantum systems operate at nonzero temperatures. For example, superconducting qubits \cite{wellard2002thermal, simbierowicz2024inherent} and quantum networks \cite{kim2022operator, cohen2017thermalization} are subject to thermal fluctuations, making it essential to design coherence-preserving techniques that remain effective in thermal environments. While frequency modulation has demonstrated efficacy in preserving coherence at zero temperature, its performance under thermal noise remains to be further investigated.

In this work, we explore the dynamics of a frequency-modulated qubit interacting with a thermal, phase-covariant reservoir, encompassing both dissipative and dephasing processes. First, we analyze a scenario where the qubit couples to a thermal dissipative reservoir at temperature $T_1$. We show that frequency modulation significantly prolongs coherence even at elevated temperatures compared to the undriven case. Next, we investigate a qubit coupled to a pure-dephasing reservoir at temperature $T_2$. In this case, frequency modulation is found to be completely ineffective due to the commutation between the qubit’s self-Hamiltonian and the interaction Hamiltonian. Finally, we consider a more general scenario where the qubit interacts independently with both a thermal dissipative and a thermal dephasing reservoir. Our results reveal that frequency modulation can still protect quantum coherence, but only if the dephasing coupling strength $\alpha$ remains below a certain threshold. We also identify the critical value of $\alpha$ beyond which frequency modulation no longer offers protection.

The paper is organized as follows. In Sec.~\ref{Model} we present a general model to describe the dynamics of a frequency-modulated qubit undergoing phase-covariant noise, using a time-local master equation. Sec.~\ref{Driven qubit interacting with a Thermal Dissipative reservoir} and Sec.~\ref{DRIVEN QUBIT INTERACTING WITH
INDEPENDENT THERMAL DISSIPATIVE AND THERMAL
DEPHASING RESERVOIRS} report the results that demonstrate the effectiveness of FM in preserving quantum coherence of the  qubit under different noisy conditions. Finally, we outline main conclusions and prospects in Sec.~\ref{CONCLUSION}.

\begin{figure*}[t]
    \centering
\includegraphics[width=0.8\textwidth]{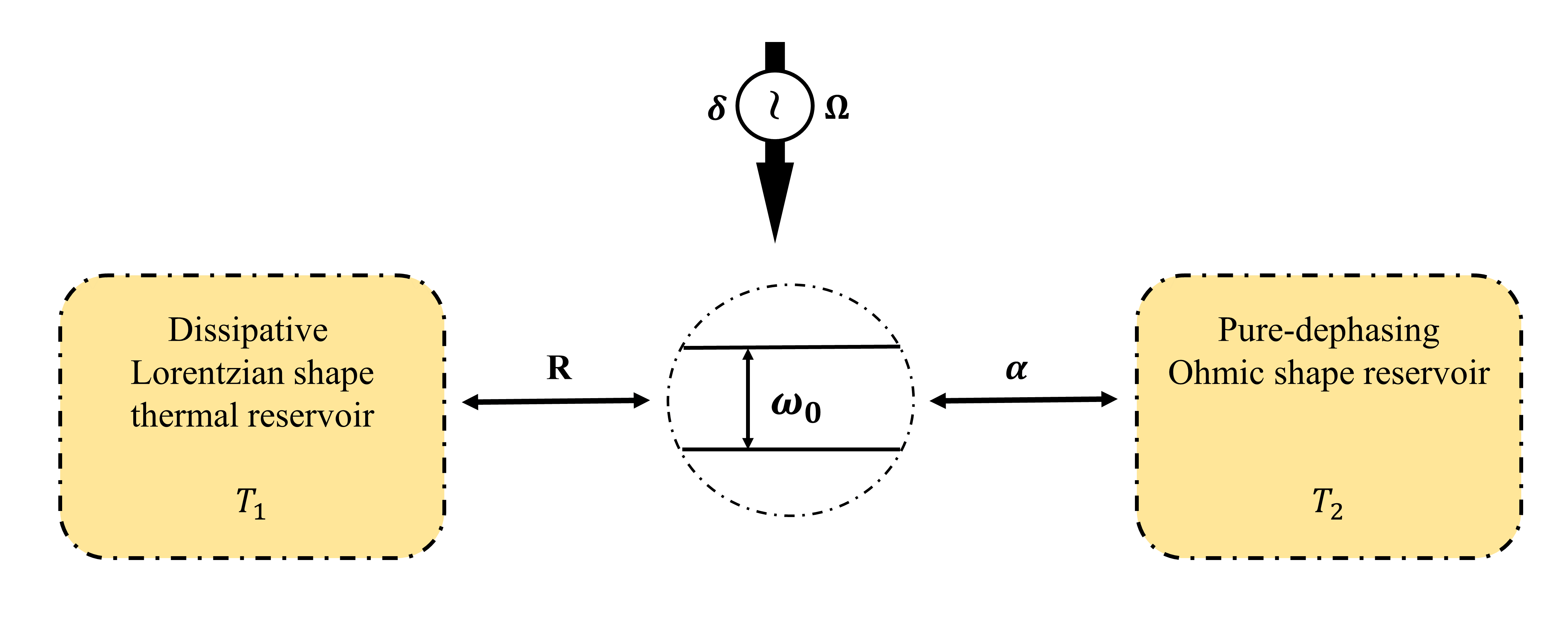}
    \caption{A sketch of the driven qubit system. A qubit (two-level atom) interacts with two independent types of ideal noise: dissipative noise at temperature $T_1$ and phase
noise at temperature $T_2$. The transition frequency of the qubit $\omega_0$ is modulated sinusoidally by an externally applied field, characterized by a modulation amplitude $\delta$ and a modulation frequency $\Omega$. $R$ and $\alpha$ represent the dimensionless coupling strengths between the qubit and the dissipative and dephasing reservoirs, respectively.}
    \label{fig:top-image}
\end{figure*}

\section{Model}
\label{Model}

The system under consideration is a qubit (two-level system), with an excited state $\ket{e}$ and a ground state $\ket{g}$. The qubit transition frequency $\omega_0$ is modulated sinusoidally by an external driving field, characterized by a modulation amplitude $\delta$ and a modulation frequency $\Omega$ ($\delta, \Omega \ll \omega_0$). The qubit interacts with two independent types of  ideal noise: amplitude damping noise and phase noise at temperature $T_1$ and $T_2$, respectively, as depicted in Fig.~\ref{fig:top-image}. The term ideal means that the noise is sufficiently weak and random \cite{yu2006quantum}. Each noise source can be modeled as a reservoir composed of an infinite array of random, broad-spectrum harmonic oscillators \cite{breuer2002theory}. The total Hamiltonian for a driven qubit coupled to two local reservoirs (noise sources) can be written as
\begin{equation}\label{Htotal}
    H = H_\mathrm{q} + H_\mathrm{r1} + H_\mathrm{r2} + H_\mathrm{I,dis} + H_\mathrm{I,deph}, 
\end{equation}
where $H_\mathrm{q}$ is the Hamiltonian of the driven qubit ($\hbar = 1$) 
\begin{equation}\label{eq: freeH}
     H_\mathrm{q} = \frac{1}{2}(\omega_0+\delta \cos({\Omega t)})  \sigma_z,
\end{equation}
with ${\sigma }_{z}$ denoting the Pauli operator along the z-direction; $H_\mathrm{r1} = \sum_{k}\omega_k a^\dagger_k a_k$ and $H_\mathrm{r2} = \sum_{l}\mu_l b^\dagger_l b_l$ are the Hamiltonians of the reservoirs, where, $a^{\dagger}_{k}, b^{\dagger}_{l}$ ($a_{k}, b_{l}$) are the creation (annihilation) operators of the damping reservoir mode $k$ with frequency $\omega_k$ and dephasing reservoir mode $l$ with frequency $\mu_l$, respectively.
$H_\mathrm{I, dis}$ ($H_\mathrm{I, deph}$) describes the interaction between the qubit and the dissipative (dephasing) reservoir modes
\begin{equation}\label{eq: H_dis}
    H_\mathrm{I,dis} = \sum_{k} \left( g_{k} \sigma_{+} a_{k} + g_{k}^{\ast} a_{k}^{\dagger} \sigma_{-} \right),
\end{equation}
\begin{equation}\label{eq: H_dip}
    H_\mathrm{I,deph} = \sum_{l} \sigma_{z} \left( f^{\ast}_{l} b^{\dagger}_{l} + f_{l} b_{l} \right),
\end{equation}
where $g_k$ ($f_l$) are the coupling constant between the qubit and mode $k$ ($l$), while the operators ${\sigma }_{\pm} = {\sigma }_{x} \pm i{\sigma }_{y}$ are the qubit raising and lowering
operators. 

The time-local master equation for this qubit, derived after tracing over the two reservoirs and considering phase-covariant noise in the interaction picture, is expressed as \cite{lankinen2016erratum, jahromi2020witnessing}
\begin{equation}
\label{eq: General MSE}
\begin{split}
    {\dot{\rho }}_{q}(t) &=-\,i{\omega }(t)[{\sigma }_{z},{\rho }_{q}(t)] 
+ \frac{\gamma_1(t)}{2}L_{1}[\rho_{q}] \\
&+\frac{\gamma_2(t)}{2}L_{2}[\rho_{q}] 
+ \frac{\gamma_3(t)}{2}L_{3}[\rho_{q}],
\end{split}
\end{equation}
where ${\rm{\omega }}(t)=\frac{1}{2}(\omega_0+\delta \cos{(\Omega t)})$ is the modulated transition frequency \cite{mortezapour2018protecting}, and
$\gamma_i(t)$ $ (i= 1,2,3)$ are the time-dependent rates associated with the Lindbladian superoperators $L_i [\rho_q]$. These superoperators are defined as \cite{breuer2016colloquium, nielsen2010quantum, he2019non} 
\begin{equation}
\label{Lindbladian superoperators}
\begin{aligned}
L_{1}[\rho_{q}] &= {\sigma }_{-}{\rho }_{q}(t){\sigma }_{+} -\,\frac{1}{2}{\sigma }_{+}{\sigma }_{-}{\rho }_{q}(t)-\,\frac{1}{2}{\rho }_{q}(t){\sigma }_{+}{\sigma }_{-}, \\
L_{2}[\rho_{q}] &= {\sigma }_{+}{\rho }_{q}(t){\sigma }_{-} -\,\frac{1}{2}{\sigma }_{-}{\sigma }_{+}{\rho }_{q}(t)-\,\frac{1}{2}{\rho }_{q}(t){\sigma }_{-}{\sigma }_{+}, \\ 
L_{3}[\rho_{q}] &= {\sigma }_{z}{\rho }_{q}(t){\sigma }_{z}-\rho_q,
\end{aligned}
\end{equation}
The three superoperators $L_1, L_2$, and $L_3$, describe dissipation, heating, 
and dephasing, respectively. The time-local master equation for this qubit combines the effects of a pure-dephasing reservoir when $\gamma_1(t) = \gamma_2(t) = 0$ and a dissipative
reservoir when $\gamma_3(t) = 0$ \cite{breuer2016colloquium, nielsen2010quantum, he2019non}. The bath temperatures are meant to be included in the general time-dependent noise rates $\gamma_i(t)$, as we shall report in the next section. 

The solution of this master equation, expressed in the basis $\{\ket{e}, \ket{g}\}$, is represented as the qubit density matrix
\begin{equation}
\label{density matrix}
 \rho_\mathrm{q}(t) =
  \begin{pmatrix}
    1-P_g(t) & \zeta(t) \\
    \zeta^*(t) & P_g(t) \\
  \end{pmatrix},
\end{equation}
where 
$\zeta(t)=\langle g\left|\rho_\mathrm{q}(t)\right|e\rangle$, i.e., the off-diagonal element of $\rho_\mathrm{q}(t)$), corresponds to the well-known $l_1$-norm of coherence \cite{baumgratz2014quantifying,streltsov2017colloquium}, and $P_g(t)=\langle g\left|\rho_{q}(t)\right|g\rangle$ represents the population of the qubit ground state. These time-dependent elements of the qubit density matrix are given by \cite{lankinen2016erratum}
\begin{equation}
\label{eq: elements}
\begin{aligned}
\zeta(t) &= \zeta(0) e^{i\Tilde{\omega}(t)-\Gamma(t)/2-\Tilde{\Gamma}(t)}, \\
P_g(t) &= e^{-\Gamma(t)}[P_g(0) + G(t)],
\end{aligned}
\end{equation}
where
\begin{equation}
\label{eq: time-dependent functions}
\begin{aligned}
\Gamma(t) &=\frac{1}{2} \int_{0}^{t} dt^\prime (\gamma_1(t^\prime) + \gamma_2(t^\prime)), \\
\Tilde{\Gamma}(t) &= \int_{0}^{t} dt^\prime \gamma_3(t^\prime), \\
\Tilde{\omega}(t) &= \int_0^t dt^\prime \omega(t^\prime), \\
G(t) &= \frac{1}{2}\int_0^t dt^\prime e^{\Gamma(t^\prime)} \gamma_1(t^\prime).
\end{aligned}
\end{equation}
To capture the dynamics of the system, it is necessary to find the time-dependent functions in Eq.~\eqref{eq: time-dependent functions}, which are associated with the time-dependent rates of the master equation. To achieve this, we employ an intuitive heuristic model that ensures complete positivity and trace preservation (CPTP) \cite{lankinen2016complete}. In this approach, the decay rates are first determined for the zero-temperature case, which are then extended to the nonzero-temperature scenario.  

In the following, we first consider a thermal dissipative reservoir and successively include a second thermal pure-dephasing reservoir.

\section{Driven qubit in a thermal dissipative reservoir}
\label{Driven qubit interacting with a Thermal Dissipative reservoir}

We first consider the case where the qubit interacts only with a dissipative reservoir. Therefore, the rate associated with pure-dephasing is zero, i.e., $\gamma_3(t) = 0$ in the master equation of Eq.~\eqref{eq: General MSE}. As discussed above, we begin with the scenario of a zero-temperature ($T_1=0$) dissipative reservoir, which further implies $\gamma_2(t) = 0$ in Eq.~\eqref{eq: General MSE}. From the Hamiltonian perspective, these conditions allow us to omit $H_\mathrm{r2}$ and $H_\mathrm{I,deph}$ in Eq.~(\ref{Htotal}). Under these assumptions, the aim is to derive the expression for the decay rate $\gamma_1(t)$. 

With the simplifications above, we can obtain the effective Hamiltonian in the interaction picture via $H_\mathrm{eff} =U^{\dagger}_0 H U_{0}+i(\partial U^{\dagger}_0 /\partial t)U_{0}$, where the unitary transformation is given by \cite{mortezapour2018protecting}
\begin{equation}\small \label{eq: unitary transformation}
U_{0}=\exp \left[-i\left\{\sum _{k}\omega _{k} a_{k}^{\dagger} a_{k}t + [\omega _{0}t +(\delta /\Omega )\sin (\Omega t)]\sigma _{z} /2\right\}\right]. 
\end{equation}
Using this transformation, the effective Hamiltonian becomes
\begin{eqnarray}\label{eq: effective Hamiltonian}
H_\mathrm{eff} &=& \sum _{k}g_{k} \sigma _{+} a_{k} e^{-i(\omega _{k} -\omega _{0} )t} e^{i(\delta /\Omega )\sin (\Omega t)} \nonumber \\
&+& \sum _{k} g_{k}^{*} a_{k}^{ \dagger } \sigma _{-} e^{i(\omega _{k} -\omega _{0} )t} e^{-i(\delta /\Omega )\sin (\Omega t)}\ .
\end{eqnarray} 
Using the Jacobi-Anger expansion, the exponential factors in Eq.~(\ref{eq: effective Hamiltonian}) can be written as
\begin{equation}
\label{Jacobi-Anger expansion}
    {e}^{\pm i(\delta /{\rm{\Omega }})\sin ({\rm{\Omega }}t)}={J}_{0}\left(\frac{\delta }{{\rm{\Omega }}}\right)+2\,\sum _{n=1}^{\infty }\,{(\pm i)}^{n}{J}_{n}\left(\frac{\delta }{{\rm{\Omega }}}\right)\,\cos (n{\rm{\Omega }}t),
\end{equation}
where ${J}_{n}\left(\frac{\delta }{{\rm{\Omega }}}\right)$ is the $n$-th Bessel function of the first kind. 

A zero-temperature reservoir implies that at most one excitation exists due to the spontaneous emission of the qubit. Consequently, the combined state of the qubit and the reservoir can be expressed as $\left| \Psi (0) \right\rangle =\frac{1}{\sqrt{2}}({ \left| e \right\rangle +\left| g \right\rangle}) \left| 0 \right\rangle$, where $\left| 0 \right\rangle$ is the vacuum state of the reservoir. The conservation of the total number of excitations leads to the time-evolved state of the system ${\left| \Psi (t) \right\rangle} =\frac{1}{\sqrt{2}} C(t) {\left| e \right\rangle} {\left| 0 \right\rangle} +\frac{1}{\sqrt{2}} {\left| g \right\rangle} {\left| 0 \right\rangle}\nonumber
+{\sum _{k}C_{k} (t){\left| g \right\rangle} {\left| 1_{k}  \right\rangle}}$, 
where ${\left| 1_{k}  \right\rangle}$ describes the presence of a single photon in mode $k$ of the reservoir, and $C_{k} (t)$ represents the corresponding probability amplitude. Solving the time-dependent Schrödinger equation for the total system yields the  differential equation for the probability amplitude $C(t)$, given by
\begin{equation}\label{eq: firtDE} 
\dot{C}(t)+\int _{0}^{t}dt^\prime  F(t,t^\prime )C(t^\prime)=0, 
\end{equation}
where the correlation function $F(t,t^\prime )$ can be expressed in terms of the continuous spectrum of the environment frequencies, as detailed in \cite{mortezapour2018protecting}, as
\begin{eqnarray}\label{eq: correlation function}
F(t,t^\prime )=\exp[i(\delta /\Omega)\{\sin(\Omega t) -\sin(\Omega t^\prime)\}] \nonumber \\
\times\int _{0}^{\infty}J(\omega)e^{-i(\omega-\omega_{0})(t-t^\prime)}d\omega, 
\end{eqnarray}
where $J(\omega)$ is the spectral density of the reservoir. We assume that the spectral density of the dissipative reservoir follows a Lorentzian profile \cite{breuer2002theory, lofranco2013dynamics}, given by 
\begin{equation}\label{eq: Lorentzian} 
J(\omega)=\frac{1}{2\pi } \frac{\gamma \lambda ^{2} }{[(\omega_{0} -\omega)^{2} +\lambda ^{2}]}, 
\end{equation}
where $\lambda$ represents the spectral width of the coupling, and $\gamma$ denotes the spontaneous emission decay rate \cite{mortezapour2018coherence}. For the analysis, we can introduce the dimensionless parameter $R =\gamma/\lambda$ which quantifies the coupling strength between the driven qubit and the dissipative reservoir (see Fig.~\ref{fig:top-image}): $R<1$ means a weak coupling regime, while $R>1$ indicates a strong coupling regime. 

With the Lorentzian spectral density of Eq.~(\ref{eq: Lorentzian}), one obtains an explicit expression for the correlation function as $F(t,t^\prime )=\frac{\gamma \lambda}{2}e^{-\lambda\left(t-t^\prime\right)}e^{i(\delta /\Omega)\left(\sin(\Omega t) -\sin(\Omega t^\prime)\right)}$, which substituted in Eq.~\eqref{eq: firtDE} gives \cite{mortezapour2018protecting}
\begin{equation}
\label{eq: probability amplitude}
\begin{aligned}
{\dot{C}}(t) &= -\,\frac{\gamma \lambda }{2} e^{i(\delta /{\Omega })\sin ({\Omega }t)} \\
&\quad \times \int_{0}^{t} dt^{\prime} \, e^{-i(\delta /{\Omega })\sin ({\Omega }t^{\prime} )} 
e^{-\lambda (t-t^{\prime} )} C(t^{\prime} ).
\end{aligned}
\end{equation}
The solution of the above equation for the probability amplitude $C(t)$ is used to obtain the quantum coherence $ \zeta(t)$ of the qubit interacting with a zero-temperature dissipative reservoir, as reported in Ref.~\cite{mortezapour2018protecting}.

After recalling the result for the zero-temperature case, we now extend the analysis to nonzero temperatures by relating the probability amplitude $C(t)$ to the decay rates $\gamma_1(t)$ and $\gamma_2(t)$ in the master equation of Eq.~\eqref{eq: General MSE}. The reservoir at a finite temperature $T_1$ introduces thermal effects, described by the Bose-Einstein mean occupation number, $\bar{n} = (e^{(\omega / K_B T_1)} - 1)^{-1}$. Here, $K_B$ represents Boltzmann’s constant \cite{scully1999quantum}, and $\omega_T = K_B T_1$ defines the thermal frequency \cite{reina2002decoherence}. The decay rates incorporating the effects of nonzero temperatures are given by  $\gamma_1(t)/2 = (\bar{n}+1)f(t)$ and $\gamma_2(t)/2 = \bar{n}f(t)$ \cite{lankinen2016erratum}, where $f(t)$ is
\begin{equation}\label{eq: f(t)}
    f(t) = -2 {\rm{Re}}\left(\frac{\dot{C(t)}}{C(t)}\right). 
\end{equation}
The time-dependent decay rate $\Gamma(t)$ is analytically calculated by substituting the above expression in Eq.~\eqref{eq: time-dependent functions}, that is
\begin{equation}
\label{Gamma-dissipative}
    \Gamma(t) = -\mathrm{ln}\left|\frac{C(t)}{C(0)}\right|^{2\Bar{n}+1}.
\end{equation}
Thanks to these expressions, from Eqs.~(\ref{eq: elements}) and (\ref{eq: time-dependent functions}) we obtain the quantum coherence $\zeta(t)$ and the population of ground state $P_g(t)$ as
\begin{equation}
\label{eq: quantum coherence-population of ground state}
\begin{aligned}
\zeta(t)=&\zeta(0)\left|\frac{C(t)}{C(0)}\right|^{2\Bar{n}+1}
\\
P_g(t)=&P_g(0)\left|\frac{C(t)}{C(0)}\right|^{2(2\Bar{n}+1)}+\frac{\Bar{n}+1}{2\Bar{n}+1}\Bigg(1-\left|\frac{C(t)}{C(0)}\right|^{2(2\Bar{n}+1)}\Bigg). 
\end{aligned}
\end{equation}
Using Eq.~\eqref{eq: quantum coherence-population of ground state}, we can now numerically compute the time behavior of qubit coherence and population, under different temperature regimes and frequency modulation settings.

\begin{figure}[!t] 
\centering
\includegraphics[width=0.48\textwidth]{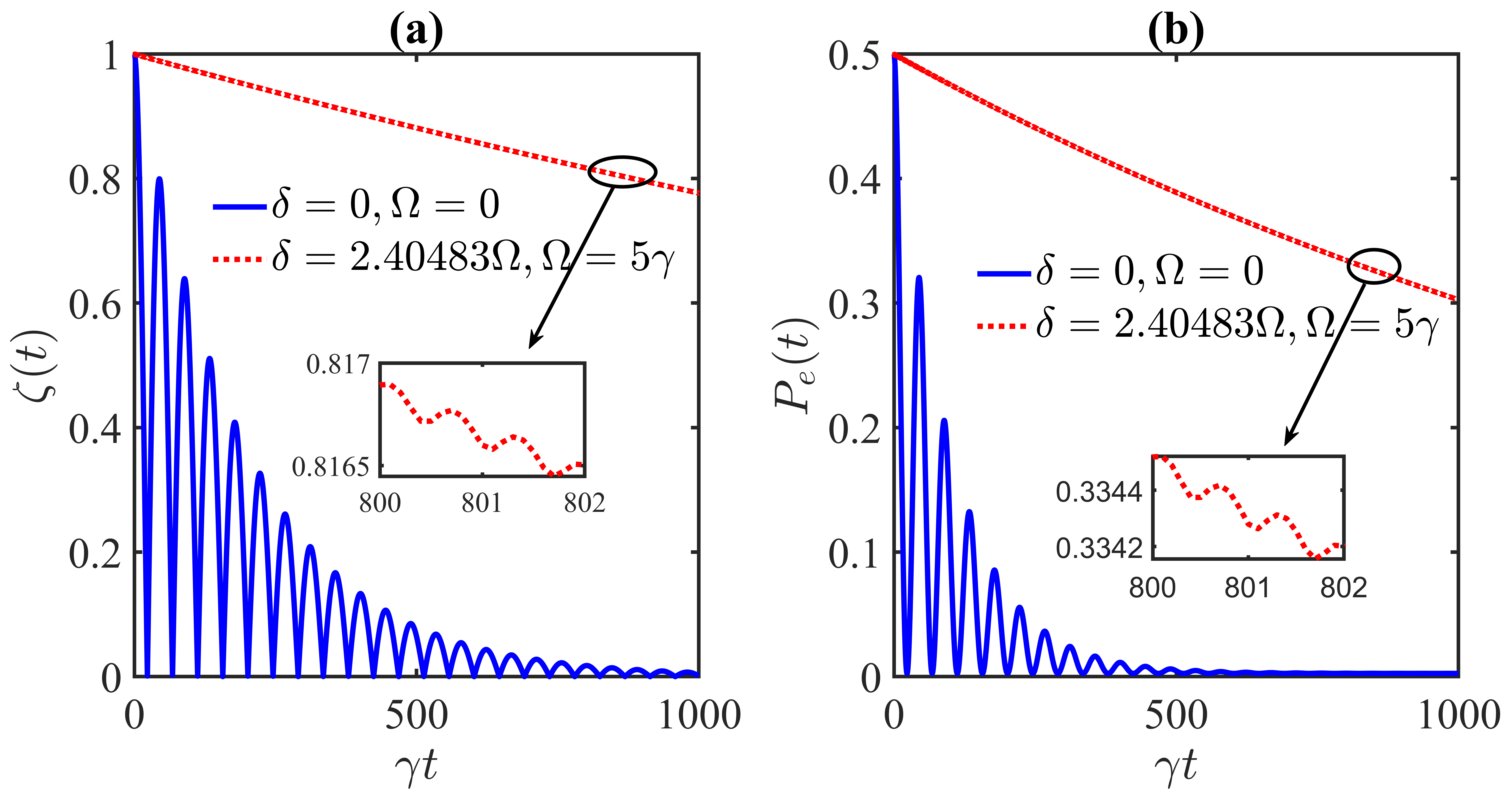}
\caption{(a) Qubit coherence $\zeta(t)$ and (b) excited state population $P_e(t)$ as functions of scaled dimensionless time $\gamma t$ under a low-temperature reservoir with $K_B T_1 = 2.6 \times 10^{-3} \hbar \omega_0$. The results are shown for optimal amplitude and frequency modulations with $\delta = 2.40483 \Omega$, $\Omega = 5 \gamma$ (dotted red line), and for the case of no driving field $\Omega=0$ (solid blue line). The qubit is in the strong coupling regime with $R = 100$.}
\label{fig-2}
\end{figure}

As a first result, Fig.~\ref{fig-2} shows the time behavior of coherence $\zeta(t)$ and excited state population $P_e(t) = 1 - P_g(t)$ as functions of the scaled (dimensionless) time $\gamma t$. We take the qubit in a strong coupling regime $R = 100$ with a low-temperature reservoir ($K_B T_1 = 2.6 \times 10^{-3} \hbar \omega_0$). Fig.~\ref{fig-2} compares two scenarios: a driven qubit with optimal modulation parameters \cite{mortezapour2018protecting} $\delta = 2.40483 \Omega$, $\Omega = 5 \gamma$ (dotted red lines) and a qubit without external driving ($\delta = 0, \Omega = 0$; solid blue lines). Notice that the optimal modulation parameters are those which enable the most effective endurance of quantumness. The plots of Fig.~\ref{fig-2} demonstrate that FM extends both coherence and excited state population significantly beyond the scenario without control. 
The non-monotonic behavior of coherence in the dotted red lines of Fig.~(\ref{fig-2}) can be associated to non-Markovian effects in the system dynamics, which are present also in the case without driving due to the chosen strong coupling conditions \cite{chanda2016delineating, teittinen2018revealing}.

\begin{figure}[!t] 
\centering
\includegraphics[width=0.48\textwidth]{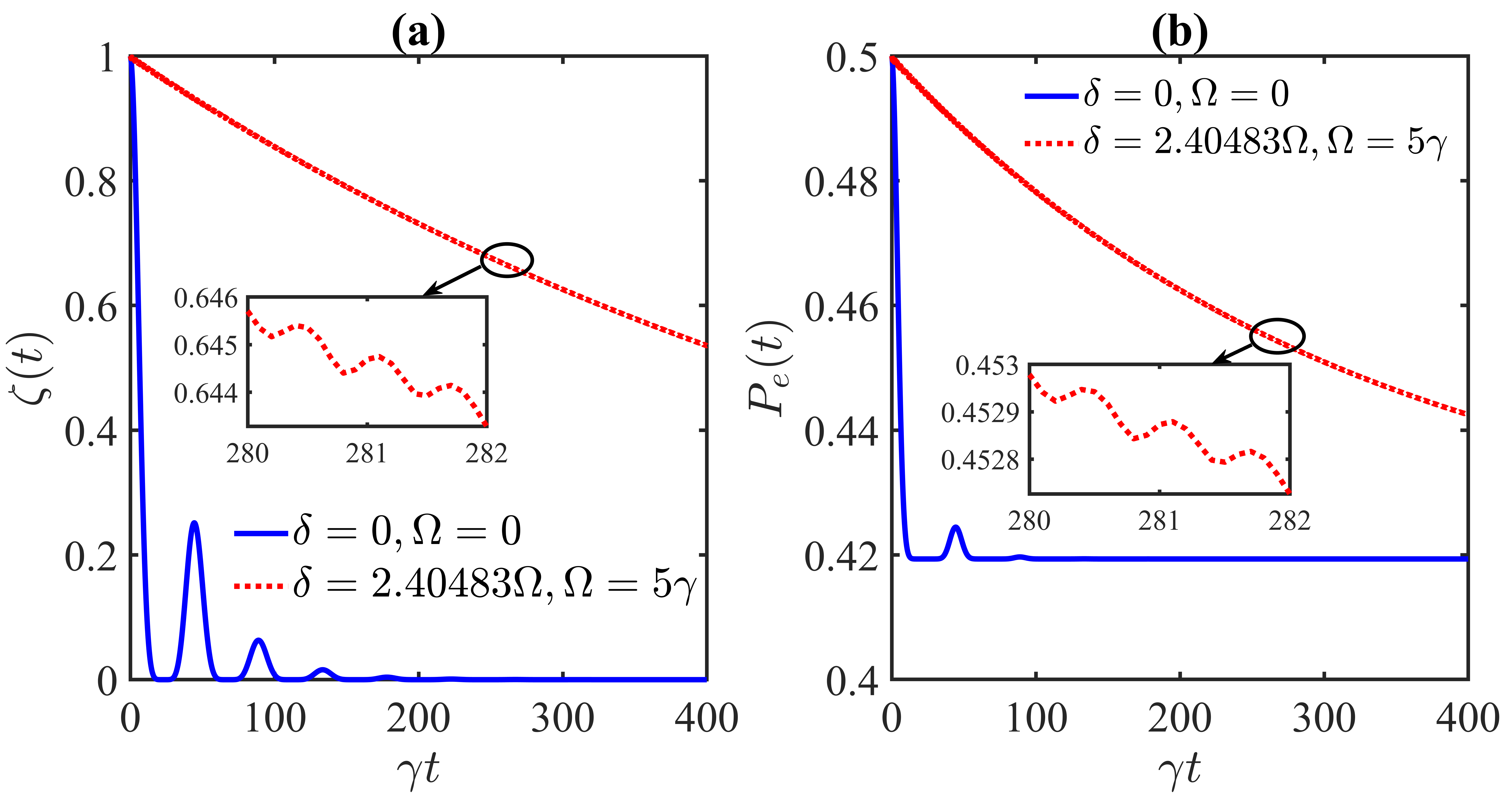}
\caption{(a) Qubit coherence $\zeta(t)$ and (b) excited state population $P_e(t)$ as functions of the scaled time $\gamma t$ under an intermediate-temperature reservoir with $K_B T_1 = 2.6 \hbar \omega_0$. The results are shown for optimal amplitude and frequency modulations with $\delta = 2.40483 \Omega$, $\Omega = 5 \gamma$ (dotted red line), and for the case of no driving field $\Omega=0$ (solid blue line). The qubit is in the strong coupling regime with $R = 100$.}
\label{fig-3}
\end{figure}

\begin{figure}[!t] 
\centering
\includegraphics[width=0.48\textwidth]{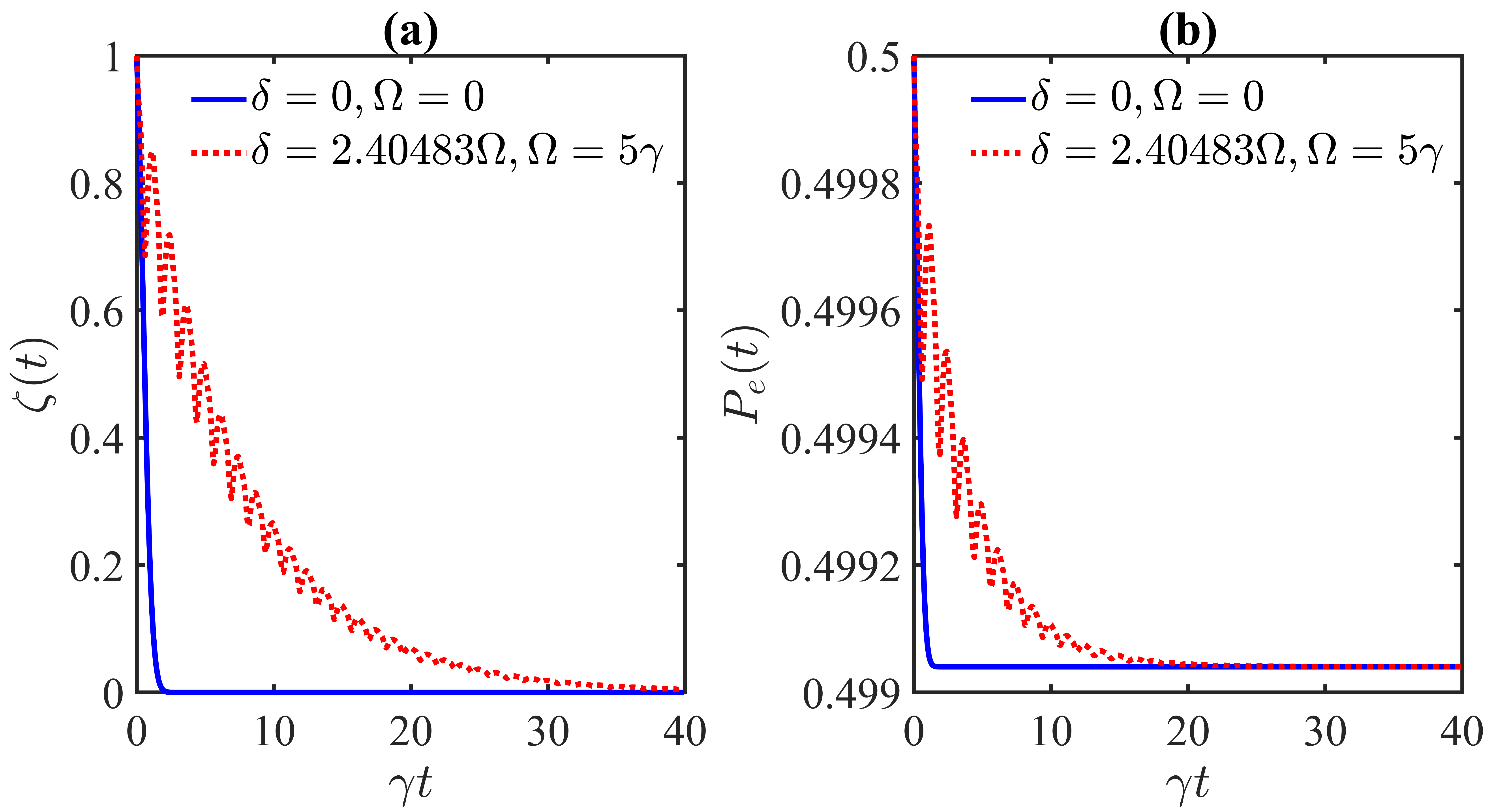}
\caption{(a) Qubit coherence $\zeta(t)$ and (b) excited state population $P_e(t)$ as functions of scaled time $\gamma t$ under a high-temperature reservoir with $K_B T_1 = 260 \hbar \omega_0$. The results are shown for optimal amplitude and frequency modulations with $\delta = 2.40483 \Omega$, $\Omega = 5 \gamma$ (dotted red line), and for the case of no driving field $\Omega=0$ (solid blue line). The qubit is in the strong coupling regime with $R = 100$.}
\label{fig-4}
\end{figure}

We then place the qubit in reservoirs at intermediate temperature ($K_B T_1 = 2.6 \hbar \omega_0$) and high temperature ($K_B T_1 = 260 \hbar \omega_0$). The plots of coherence $\zeta(t)$ and  excited state population $P_e(t)$ as functions of scaled time  are reported in Figs.~\ref{fig-3} and \ref{fig-4}, respectively. As done before, Figs.~\ref{fig-3} and \ref{fig-4} show the case without driving field ($\delta = 0, \Omega = 0$) the case with the action of a driving field under optimal modulation parameters (same as before). Again, optimal FM allow a preservation of quantum coherence and excited state population much longer than the case without control, as illustrated by the dotted red lines in Figs.~\ref{fig-3} and \ref{fig-4}. Fluctuations in coherence and population, appearing in Figs.~\ref{fig-3} and \ref{fig-4}, again indicate a signature of non-Markovianity in the system. 

These results prove that the technique of FM of the qubit effectively mitigates the dissipative effects of temperature, even in the high-temperature regime.

\section{Driven qubit under independent thermal dissipative and dephasing noises}

\label{DRIVEN QUBIT INTERACTING WITH
INDEPENDENT THERMAL DISSIPATIVE AND THERMAL
DEPHASING RESERVOIRS}

In this section, we consider a single qubit interacting with a thermal phase-covariant reservoir, made of two types of environmental noise to better approximate real-world conditions: dissipative noise at temperature $T_1$ and pure-dephasing noise at temperature $T_2$. 

We begin by only considering a pure-dephasing noise to simplify the analysis. This assumption sets the rates $\gamma_1(t) = \gamma_2(t) = 0$ in the master equation of Eq.~\eqref{eq: General MSE}. Under these conditions, $H_\mathrm{r1}$ and $H_\mathrm{I,dis}$ are neglected. The effective Hamiltonian in the interaction picture $H^{\prime}_\mathrm{eff} =U^{\dagger}_0 H U_{0}+i(\partial U^{\dagger}_0/\partial t)U_{0}$, 
with the unitary evolution operator from Eq.~\eqref{eq: unitary transformation}, is given by
\begin{eqnarray}\label{eq: effective Hamiltonian II} H^{\prime}_\mathrm{eff}&=& \sigma_{z} \sum_{l} \Bigg(f_{l} b_{l}^{\dagger} e^{i \mu_l t} + f_{l}^{*} b_{l} e^{-i \mu_l t}\Bigg).
\end{eqnarray}
As is evident from the above equation, FM does not affect the system evolution because the interaction Hamiltonian $H_\mathrm{I,deph}$ of Eq.~(\ref{eq: H_dip})  commutes with the free Hamiltonian of the qubit $H_\mathrm{q}$ of Eq.~(\ref{eq: freeH}). This is an important observation, as it shows that FM is completely ineffective in decoupling the qubit from the dephasing environment. Before studying the role of both dissipative and dephasing noise at given temperature, it is useful to briefly review the exact dynamics of the qubit interacting with a pure-dephasing reservoir \cite{palma1996quantum}.

The effective Hamiltonian in Eq.~\eqref{eq: effective Hamiltonian II} permits to derive the time-evolution operator for the qubit and the pure-dephasing reservoir as \cite{palma1996quantum} 
\begin{equation} \label{eq: time evolution operator}
U(t) = \exp\left[\frac{\sigma_z}{2} \sum_l \left(b_l^\dagger \xi_l(t) - b_l \xi_l^*(t)\right)\right], 
\end{equation} 
where $\xi_l(t) = 2 f_l (1 - e^{i \mu_l t})/\mu_l$. For clarity and without loss of generality, let us consider the qubit at $t = 0$ interacting with a zero-temperature pure-dephasing reservoir. The initial state of the system is given as $\ket{\psi(0)}=\left(c_1\ket{e} + c_2\ket{g}\right)\ket{0_l}$, where $\ket{0_l}$ represents the vacuum state of the field mode $l$. Consequently, the evolved state of both qubit and reservoir is $\ket{\psi(t)} = c_1 \ket{e}\ket{\frac{1}{2}\xi_l(t)} + c_2 \ket{g}\ket{-\frac{1}{2}\xi_l(t)}$, where $\ket{\frac{1}{2}\xi_l(t)}$ is a coherent state with amplitude $\frac{1}{2}\xi_l(t)$. Tracing over the reservoir modes reveals that the populations remain unaffected, while the off-diagonal elements of the qubit density matrix, ruled by $\zeta(t)= \zeta(0) e^{-\Tilde{\Gamma}(t)}$, decay over time. This decay arises due to the diminishing overlap between the different field states associated with the qubit evolution \cite{palma1996quantum}. 

In a more general scenario where the reservoir is at a finite temperature ($T_2 > 0$), the time dependent decay rate $\gamma_3(t)$ and dephasing factor $\Tilde{\Gamma}(t)$ become \cite{lankinen2016complete, haikka2013non}
\begin{equation}\label{eq: dephasing rate}
    \gamma_3(t) = \int_0^\infty d\omega J(\omega) \coth(\hbar \omega / 2k_BT_2) \frac{\sin(\omega t)}{\omega},
\end{equation}
\begin{equation} \label{eq: dephasing factor}
\Tilde{\Gamma}(t) = 2 \int_0^\infty d\omega J(\omega) \coth(\hbar \omega / 2k_BT_2) \frac{1 - \cos(\omega t)}{\omega^2},
\end{equation}
where $J(\omega)$ is the spectral density. Here, we assume that the dephasing reservoir has an Ohmic spectral density \cite{breuer2002theory}, given by 
\begin{equation} \label{eq: ohmic spectral density} J(\omega) = \alpha \omega^s e^{-\omega/\omega_c}, \end{equation}
characterized by a cutoff frequency $\omega_c$ that depends on the particular physical system under consideration and ensures that $J(\omega) \to 0$ for $\omega \gg \omega_c$ \cite{reina2002decoherence}. In this context, $\alpha > 0$ is a dimensionless constant that defines the strength of the system-bath coupling (see Fig.~\ref{fig:top-image}), while $s$ denotes the dimensionality of the field. By changing the parameter $s$, one can transition from sub-Ohmic reservoirs $(s < 1)$ to Ohmic reservoirs $(s = 1)$ and super-Ohmic reservoirs $(s > 1)$, respectively. In our analysis we focus on the one-dimensional field scenario ($s = 1$). Substituting Eq.~\eqref{eq: dephasing rate} into Eq.~\eqref{eq: time-dependent functions} allows us to get the quantum coherence $\zeta(t)$ of Eq.~(\ref{eq: elements}) for the thermal pure-dephasing reservoir.

We now switch on both noises, dissipative and dephasing, at the same time. Using the additivity property of coherence decay rates \cite{yu2006quantum}, the relaxation rate of a system subjected to multiple weak noise sources is equal to the sum of the relaxation rates corresponding to each individual noise source, that is $\gamma_\mathrm{tot}=\frac{\gamma_1(t)}{2} + \frac{\gamma_2(t)}{2} + \gamma_3(t)$.
Therefore, the time-dependent quantum coherence in this general scenario can be obtained by multiplying the coherence functions found above for the dissipative and pure-dephasing dynamics separately. 

Recall that FM cannot decouple the qubit from the dephasing reservoir and the system operates in the Markovian regime for the one-dimensional field scenario ($s = 1$), when only dephasing acts. In fact, it is known that backflow of information and recoherence occur only when $s > 2$ \cite{haikka2013non}. 
So, the value of the dephasing coupling $\alpha$ is expected to be very relevant on the possibility of maintaining quantum coherence within the qubit, when analyzing the evolution of coherence under both noises across a range of temperatures. 

\begin{figure}[!t] 
\centering
\includegraphics[width=0.25\textwidth]{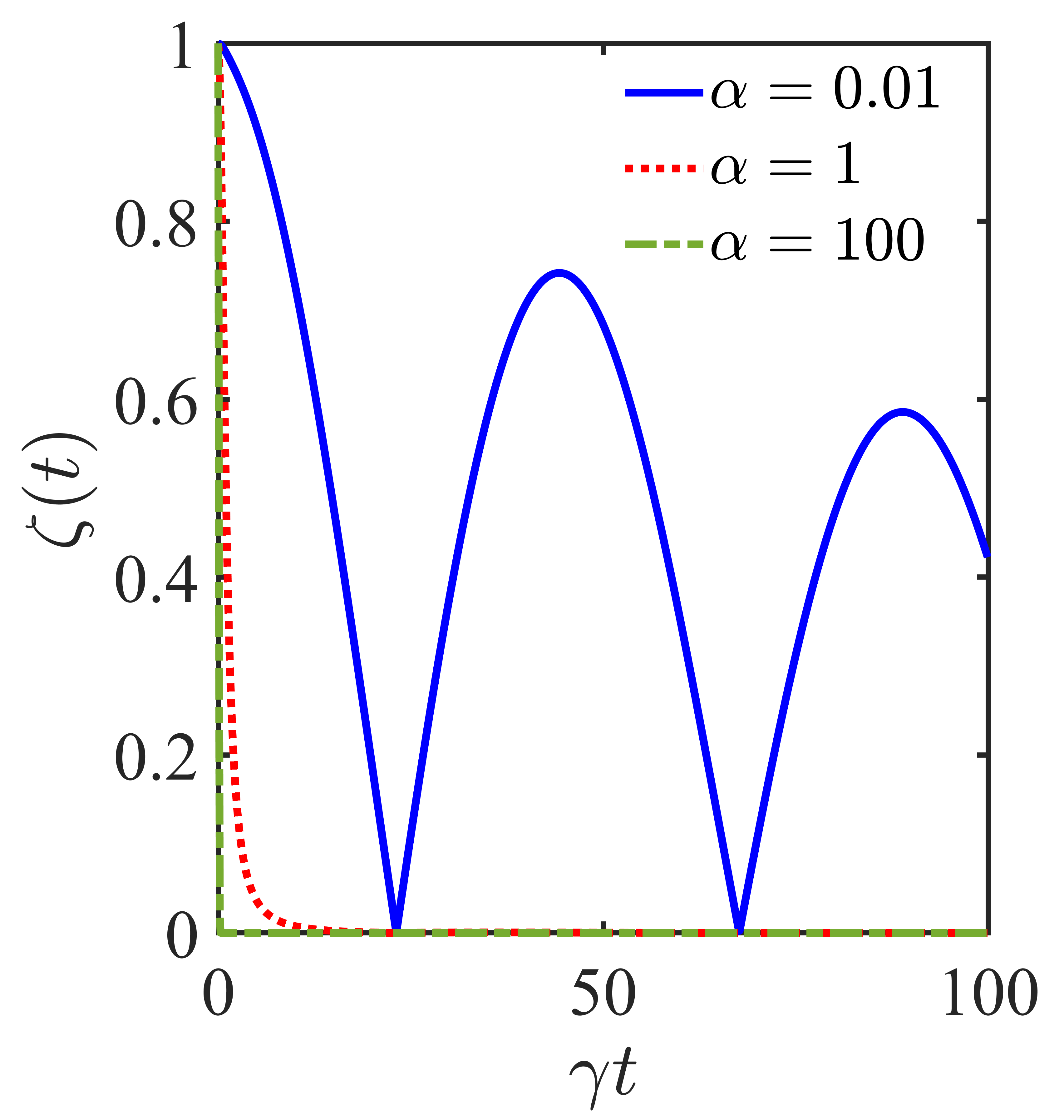}
\caption{Qubit coherence $\zeta(t)$ as a function of scaled time $\gamma t$ for different values of the dephasing coupling $\alpha$. The values of other parameters are: $R=100$, $\delta = 0$, $\Omega = 0$. Both dissipative and dephasing reservoirs are set to zero temperature ($T_1 = T_2 = 0$).}
\label{fig-5}
\end{figure}

The above consideration leads us to firstly fix the temperatures of the reservoirs equal to zero and exclude control effects from the analysis, in order to identify a typical value of $\alpha$ which is not very detrimental for qubit coherence when time goes by. Moreover, we choose $R = 100$ for the dissipative environment, so to have strong coupling and non-Markovian dynamics stemming from  system-environment excitation exchange (as noticed in Sec.~\ref{Driven qubit interacting with a Thermal Dissipative reservoir}). From Fig.~\ref{fig-5}, we observe that weak coupling strengths with the dephasing reservoir, of the order of $\alpha=0.01$, are already enough to permit coherence to last orders of magnitudes longer than the case of stronger couplings. 

\begin{figure}[!t] 
\centering
\includegraphics[width=0.48\textwidth]{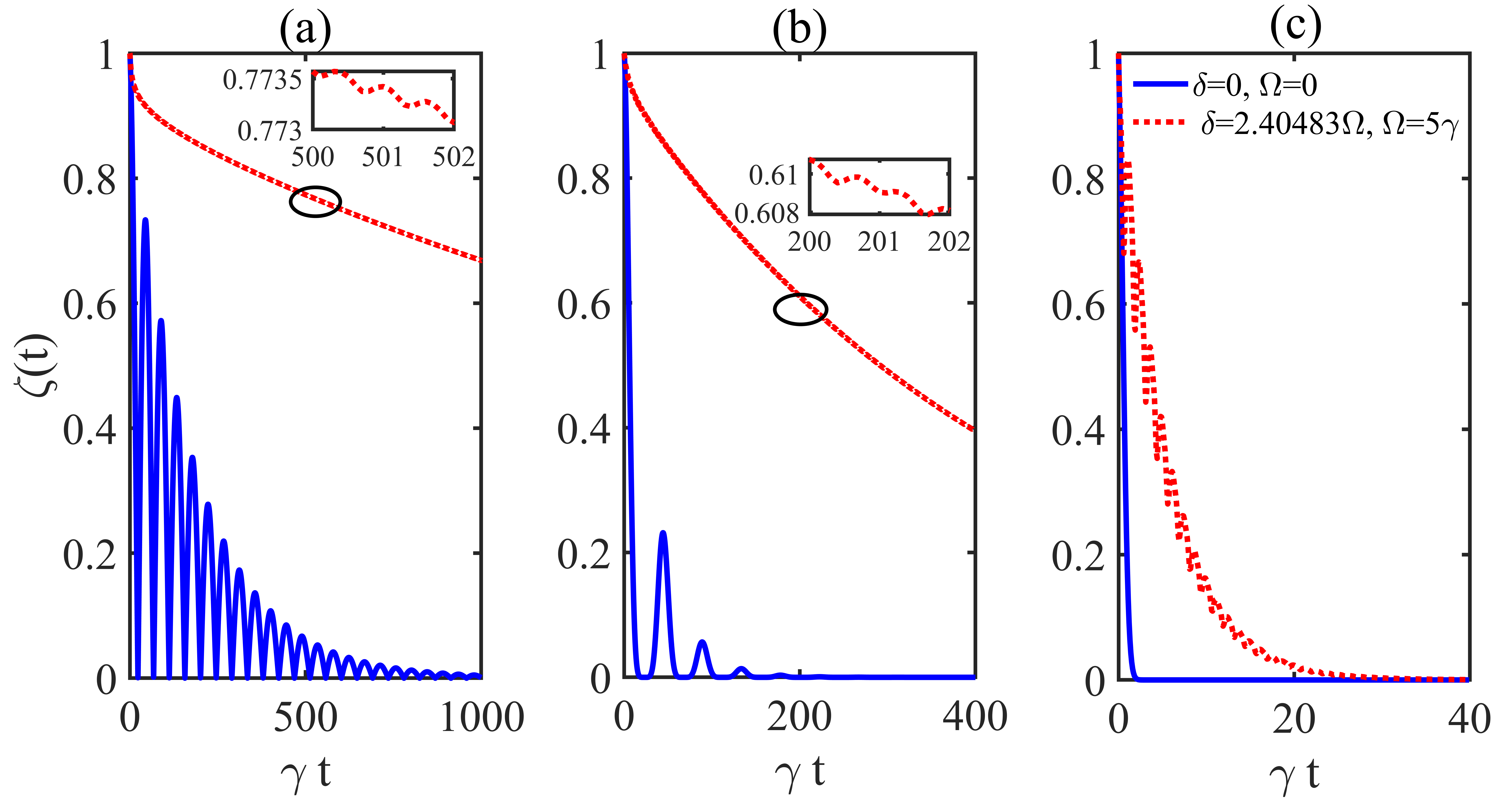}
\caption{Qubit coherence $\zeta(t)$ as a function of scaled time $\gamma t$ for different temperature regimes: (a) $K_B T_1 = 2.6 \times 10^{-3} \hbar \omega_0 $, $K_B T_2 = 10^{-5} \hbar \omega_c$, (b) $K_B T_1 = 2.6 \hbar \omega_0$, $K_B T_2 = 10^{-2} \hbar \omega_c$, (c)  $K_B T_1 = 2.6 \times 10^{2} \hbar \omega_0$, $K_B T_2 = \hbar \omega_c$. The results are shown for optimal amplitude and frequency modulations with $\delta = 2.40483 \Omega$, $\Omega = 5 \gamma$ (dotted red line), and no diving $\Omega=0$ (solid blue line). The qubit is in the weak coupling regime with the pure-dephasing reservoir ($\alpha = 0.01$) and in the strong coupling regime with the dissipative reservoir ($R = 100$).}
\label{fig-6}
\end{figure}

Therefore, we set $\alpha = 0.01$ and $R = 100$ to numerically study the time behavior of qubit coherence at different temperature regimes. Fig.~\ref{fig-6} presents the evolution of coherence $\zeta(t)$ as a function of the scaled time $\gamma t$, for a qubit experiencing a range of temperatures going from low to high values. As done for Figs.~\ref{fig-2}-\ref{fig-4}, here we compare the results for a driven qubit with optimal modulation parameters ($\delta = 2.40483 \Omega$ and $\Omega = 5 \gamma$, dotted red lines) and those for a qubit without control ($\delta = 0$ and $\Omega = 0$, solid blue lines).

From Fig.~\ref{fig-6}, we clearly see that applying FM to the qubit preserves coherence for a much longer time compared to the case without control. Importantly, coherence fluctuates well above zero during its evolution for the suitably frequency-modulated qubit (dotted red lines), so avoiding the zeros occurring during the oscillations in the case without control (blue solid lines). Fluctuations indicate a signature of non-Markovianity in the qubit, which is present even in the high-temperature regime due to the interplay between FM and strong coupling with the dissipative noise. The results indicate that FM has the capability to efficiently preserve quantum coherence within the qubit when the dephasing coupling $\alpha$ is sufficiently small (of the order of $0.01$). 

An interesting question now arises: what is the threshold value $\alpha_\mathrm{th}$ of the dephasing coupling beyond which the enhancement effect due to qubit FM no longer occurs, such that coherence evolves equally for a controlled and uncontrolled qubit?

To address this question, the threshold value $\alpha_{\mathrm{th}}$ in our simulations is defined as the value of $\alpha$ at which, when plotting the coherence as a function of time for the two cases, namely without modulation ($d = 0$) and with modulation ($d = 2.40483\,\Omega$), their coherence times become equal.  

In Markovian open quantum systems, the coherence time $t_c$ is defined as the time when $\zeta(t_c) = \zeta(0)/e$, i.e.,
\begin{equation}
t_c = \frac{1}{\gamma}.
\end{equation}
However, for a non-Markovian system, this simple exponential model does not hold, and defining the coherence time is more challenging. Due to the non-Markovian nature of the dynamics, coherence plots exhibit fluctuations. To overcome this, we extract the envelope of each oscillating curve, which mimics the monotonic decay characteristic of Markovian behavior. Using this envelope, we can define the coherence time in a way consistent with the Markovian approach described above.  

Applying this definition, at very low temperature (case of Fig.~\ref{fig-6}a) we obtain $\alpha_{\mathrm{th}} = 0.75$. Consequently, FM-based quantumness enhancement at low temperature occurs when $\alpha \leq 0.75$. At intermediate temperature (case of Fig.~\ref{fig-6}b), we find $\alpha_{\mathrm{th}} = 0.6$. Furthermore, our numerical analysis establishes that $\alpha_{\mathrm{th}}$ decreases monotonically as temperature increases, reaching its minimum value of $0.45$ in the infinite-temperature limit (case of Fig.~\ref{fig-6}c).
Determining these thresholds is crucial because they identify the boundaries where FM ceases to be effective to genuinely enhance coherence. They serve as design guidelines for optimizing control parameters $\Gamma$ and $\delta$, ensuring that resources are used efficiently.

%Numerical analysis shows that at very low temperatures (case of Fig.~\ref{fig-6}a) $\alpha_\mathrm{th}= 0.25$. Consequently, FM-based quantumness enhancement at low temperatures happens when $\alpha \leq 0.25$.  
%At higher temperatures (case of Fig.~\ref{fig-6}b), we find that $\alpha_\mathrm{th}=0.15$.
%Has this threshold value of $\alpha$ been found for the case of Fig. 6b or Fig. 6c? I assumed Fig. 6b, where the protection is more relevant. However, we then write that 0.15 is the minium values of $\alpha$ reached at infinite temperature. If $\alpha_\mathrm{th}$ decreases monotonically when temperature increases, the values cannot coincide. So, what is $\alpha_\mathrm{th}$ for Fig. 6b, for Fig. 6c, and for infinite temperature? Please, check this carefully.
%We have also numerically established that $\alpha_\mathrm{th}$ decreases monotonically as the temperature increases. Interestingly, this trend continues until $\alpha_\mathrm{th}$ reaches its minimum value of 0.15 at infinite temperature. 

\section{Conclusion}\label{CONCLUSION}

In this work, we have explored the impact of the FM control technique in preserving quantum coherence within the thermal noisy system by mitigating the detrimental effect of temperature. Through a meticulous analysis of the coherence mechanisms and a focus on relevant time scales, we showed that our approach effectively enhances coherence time in the quantum system subjected to thermal noise. We first considered the general time-local master equation for a driven qubit subjected to phase-covariant noise. Then, we considered two scenarios: the interaction of a modulated qubit with (i) a thermal dissipative reservoir and (ii) both thermal dissipative and dephasing reservoirs. In the first case, we established that FM effectively prolongs coherence within the system. In the second scenario, we initially examined the interaction of a modulated qubit with a pure-dephasing reservoir and demonstrated that the FM technique has no impact on decoupling the system from the dephasing reservoir. Moreover, we showed that when the modulated qubit interacts simultaneously with two local reservoirs—a thermal dissipative reservoir and a pure-dephasing reservoir—FM is a simple yet effective technique for mitigating the adverse effects of temperature, but only when the dephasing coupling strength $\alpha$ is sufficiently small. However, we determined a threshold for $\alpha$, beyond which the enhancement effect no longer occurs. Furthermore, we have numerically shown that $\alpha_{th}$ decreases consistently as the temperature increases.

FM can be employed as a simple yet effective technique can be employed to shield the quantum system against decoherence. This technique has already been successfully utilized to fabricate and control superconducting quantum circuits with Josephson junctions \cite{silveri2017quantum, tuorila2010stark, nakamura2001rabi, li2013motional, oliver2005mach}, and semiconductor artificial atoms \cite{cao2013ultrafast}. These successful applications reassure the audience about the effectiveness of FM, instilling a sense of confidence in its potential \cite{trabesinger2017quantum}.

\begin{acknowledgments}
R.L.F. acknowledges support by MUR (Ministero dell’Università e della Ricerca) through the following projects: PNRR Project ICON-Q – Partenariato Esteso NQSTI – PE00000023 – Spoke 2 – CUP: J13C22000680006, PNRR Project QUANTIP – Partenariato Esteso NQSTI – PE00000023 – Spoke 9 – CUP: E63C22002180006. V. M. acknowledge PNRR MUR project  NQSTI (Grant No. PE0000023). F.N. acknowledges support by the I+D+i project MADQuantum-CM, financed by the European Union
NextGeneration-EU, Madrid Government and by the
PRTR.
\end{acknowledgments}

%\label{secIII}

\bibliography{KhazaeiShadfar_etal_PRA}

\begin{thebibliography}{91}%
\makeatletter
\providecommand \@ifxundefined [1]{%
 \@ifx{#1\undefined}
}%
\providecommand \@ifnum [1]{%
 \ifnum #1\expandafter \@firstoftwo
 \else \expandafter \@secondoftwo
 \fi
}%
\providecommand \@ifx [1]{%
 \ifx #1\expandafter \@firstoftwo
 \else \expandafter \@secondoftwo
 \fi
}%
\providecommand \natexlab [1]{#1}%
\providecommand \enquote  [1]{``#1''}%
\providecommand \bibnamefont  [1]{#1}%
\providecommand \bibfnamefont [1]{#1}%
\providecommand \citenamefont [1]{#1}%
\providecommand \href@noop [0]{\@secondoftwo}%
\providecommand \href [0]{\begingroup \@sanitize@url \@href}%
\providecommand \@href[1]{\@@startlink{#1}\@@href}%
\providecommand \@@href[1]{\endgroup#1\@@endlink}%
\providecommand \@sanitize@url [0]{\catcode `\\12\catcode `\$12\catcode `\&12\catcode `\#12\catcode `\^12\catcode `\_12\catcode `\%12\relax}%
\providecommand \@@startlink[1]{}%
\providecommand \@@endlink[0]{}%
\providecommand \url  [0]{\begingroup\@sanitize@url \@url }%
\providecommand \@url [1]{\endgroup\@href {#1}{\urlprefix }}%
\providecommand \urlprefix  [0]{URL }%
\providecommand \Eprint [0]{\href }%
\providecommand \doibase [0]{https://doi.org/}%
\providecommand \selectlanguage [0]{\@gobble}%
\providecommand \bibinfo  [0]{\@secondoftwo}%
\providecommand \bibfield  [0]{\@secondoftwo}%
\providecommand \translation [1]{[#1]}%
\providecommand \BibitemOpen [0]{}%
\providecommand \bibitemStop [0]{}%
\providecommand \bibitemNoStop [0]{.\EOS\space}%
\providecommand \EOS [0]{\spacefactor3000\relax}%
\providecommand \BibitemShut  [1]{\csname bibitem#1\endcsname}%
\let\auto@bib@innerbib\@empty
%</preamble>
\bibitem [{\citenamefont {Leggett}(1980)}]{leggett1980suppl}%
  \BibitemOpen
  \bibfield  {author} {\bibinfo {author} {\bibfnamefont {A.}~\bibnamefont {Leggett}},\ }\bibfield  {title} {\bibinfo {title} {Suppl. prog. theor. phys.},\ }\href@noop {} {\bibfield  {journal} {\bibinfo  {journal} {Suppl. Prog. Theor. Phys}\ }\textbf {\bibinfo {volume} {69}},\ \bibinfo {pages} {80} (\bibinfo {year} {1980})}\BibitemShut {NoStop}%
\bibitem [{\citenamefont {Yao}\ \emph {et~al.}(2015)\citenamefont {Yao}, \citenamefont {Xiao}, \citenamefont {Ge},\ and\ \citenamefont {Sun}}]{yao2015quantum}%
  \BibitemOpen
  \bibfield  {author} {\bibinfo {author} {\bibfnamefont {Y.}~\bibnamefont {Yao}}, \bibinfo {author} {\bibfnamefont {X.}~\bibnamefont {Xiao}}, \bibinfo {author} {\bibfnamefont {L.}~\bibnamefont {Ge}},\ and\ \bibinfo {author} {\bibfnamefont {C.}~\bibnamefont {Sun}},\ }\bibfield  {title} {\bibinfo {title} {Quantum coherence in multipartite systems},\ }\href@noop {} {\bibfield  {journal} {\bibinfo  {journal} {Phys. Rev. A}\ }\textbf {\bibinfo {volume} {92}},\ \bibinfo {pages} {022112} (\bibinfo {year} {2015})}\BibitemShut {NoStop}%
\bibitem [{\citenamefont {Streltsov}\ \emph {et~al.}(2015)\citenamefont {Streltsov}, \citenamefont {Singh}, \citenamefont {Dhar}, \citenamefont {Bera},\ and\ \citenamefont {Adesso}}]{streltsov2015measuring}%
  \BibitemOpen
  \bibfield  {author} {\bibinfo {author} {\bibfnamefont {A.}~\bibnamefont {Streltsov}}, \bibinfo {author} {\bibfnamefont {U.}~\bibnamefont {Singh}}, \bibinfo {author} {\bibfnamefont {H.~S.}\ \bibnamefont {Dhar}}, \bibinfo {author} {\bibfnamefont {M.~N.}\ \bibnamefont {Bera}},\ and\ \bibinfo {author} {\bibfnamefont {G.}~\bibnamefont {Adesso}},\ }\bibfield  {title} {\bibinfo {title} {Measuring quantum coherence with entanglement},\ }\href@noop {} {\bibfield  {journal} {\bibinfo  {journal} {Phys. Rev. Lett.}\ }\textbf {\bibinfo {volume} {115}},\ \bibinfo {pages} {020403} (\bibinfo {year} {2015})}\BibitemShut {NoStop}%
\bibitem [{\citenamefont {Hu}\ \emph {et~al.}(2016)\citenamefont {Hu}, \citenamefont {Milne}, \citenamefont {Zhang},\ and\ \citenamefont {Fan}}]{hu2016quantum}%
  \BibitemOpen
  \bibfield  {author} {\bibinfo {author} {\bibfnamefont {X.}~\bibnamefont {Hu}}, \bibinfo {author} {\bibfnamefont {A.}~\bibnamefont {Milne}}, \bibinfo {author} {\bibfnamefont {B.}~\bibnamefont {Zhang}},\ and\ \bibinfo {author} {\bibfnamefont {H.}~\bibnamefont {Fan}},\ }\bibfield  {title} {\bibinfo {title} {Quantum coherence of steered states},\ }\href@noop {} {\bibfield  {journal} {\bibinfo  {journal} {Sci. Rep.}\ }\textbf {\bibinfo {volume} {6}},\ \bibinfo {pages} {1} (\bibinfo {year} {2016})}\BibitemShut {NoStop}%
\bibitem [{\citenamefont {Radhakrishnan}\ \emph {et~al.}(2016)\citenamefont {Radhakrishnan}, \citenamefont {Parthasarathy}, \citenamefont {Jambulingam},\ and\ \citenamefont {Byrnes}}]{radhakrishnan2016distribution}%
  \BibitemOpen
  \bibfield  {author} {\bibinfo {author} {\bibfnamefont {C.}~\bibnamefont {Radhakrishnan}}, \bibinfo {author} {\bibfnamefont {M.}~\bibnamefont {Parthasarathy}}, \bibinfo {author} {\bibfnamefont {S.}~\bibnamefont {Jambulingam}},\ and\ \bibinfo {author} {\bibfnamefont {T.}~\bibnamefont {Byrnes}},\ }\bibfield  {title} {\bibinfo {title} {Distribution of quantum coherence in multipartite systems},\ }\href@noop {} {\bibfield  {journal} {\bibinfo  {journal} {Phys. Rev. Lett.}\ }\textbf {\bibinfo {volume} {116}},\ \bibinfo {pages} {150504} (\bibinfo {year} {2016})}\BibitemShut {NoStop}%
\bibitem [{\citenamefont {Ma}\ \emph {et~al.}(2016)\citenamefont {Ma}, \citenamefont {Yadin}, \citenamefont {Girolami}, \citenamefont {Vedral},\ and\ \citenamefont {Gu}}]{ma2016converting}%
  \BibitemOpen
  \bibfield  {author} {\bibinfo {author} {\bibfnamefont {J.}~\bibnamefont {Ma}}, \bibinfo {author} {\bibfnamefont {B.}~\bibnamefont {Yadin}}, \bibinfo {author} {\bibfnamefont {D.}~\bibnamefont {Girolami}}, \bibinfo {author} {\bibfnamefont {V.}~\bibnamefont {Vedral}},\ and\ \bibinfo {author} {\bibfnamefont {M.}~\bibnamefont {Gu}},\ }\bibfield  {title} {\bibinfo {title} {Converting coherence to quantum correlations},\ }\href@noop {} {\bibfield  {journal} {\bibinfo  {journal} {Phys. Rev. Lett.}\ }\textbf {\bibinfo {volume} {116}},\ \bibinfo {pages} {160407} (\bibinfo {year} {2016})}\BibitemShut {NoStop}%
\bibitem [{\citenamefont {Malvezzi}\ \emph {et~al.}(2016)\citenamefont {Malvezzi}, \citenamefont {Karpat}, \citenamefont {{\c{C}}akmak}, \citenamefont {Fanchini}, \citenamefont {Debarba},\ and\ \citenamefont {Vianna}}]{malvezzi2016quantum}%
  \BibitemOpen
  \bibfield  {author} {\bibinfo {author} {\bibfnamefont {A.}~\bibnamefont {Malvezzi}}, \bibinfo {author} {\bibfnamefont {.~G.}\ \bibnamefont {Karpat}}, \bibinfo {author} {\bibfnamefont {B.}~\bibnamefont {{\c{C}}akmak}}, \bibinfo {author} {\bibfnamefont {F.}~\bibnamefont {Fanchini}}, \bibinfo {author} {\bibfnamefont {T.}~\bibnamefont {Debarba}},\ and\ \bibinfo {author} {\bibfnamefont {R.}~\bibnamefont {Vianna}},\ }\bibfield  {title} {\bibinfo {title} {Quantum correlations and coherence in spin-1 heisenberg chains},\ }\href@noop {} {\bibfield  {journal} {\bibinfo  {journal} {Phys. Rev. B}\ }\textbf {\bibinfo {volume} {93}},\ \bibinfo {pages} {184428} (\bibinfo {year} {2016})}\BibitemShut {NoStop}%
\bibitem [{\citenamefont {Li}\ and\ \citenamefont {Lin}(2016)}]{li2016quantum}%
  \BibitemOpen
  \bibfield  {author} {\bibinfo {author} {\bibfnamefont {Y.-C.}\ \bibnamefont {Li}}\ and\ \bibinfo {author} {\bibfnamefont {H.-Q.}\ \bibnamefont {Lin}},\ }\bibfield  {title} {\bibinfo {title} {Quantum coherence and quantum phase transitions},\ }\href@noop {} {\bibfield  {journal} {\bibinfo  {journal} {Scientific reports}\ }\textbf {\bibinfo {volume} {6}},\ \bibinfo {pages} {26365} (\bibinfo {year} {2016})}\BibitemShut {NoStop}%
\bibitem [{\citenamefont {Chen}\ \emph {et~al.}(2016)\citenamefont {Chen}, \citenamefont {Cui}, \citenamefont {Zhang},\ and\ \citenamefont {Fan}}]{chen2016coherence}%
  \BibitemOpen
  \bibfield  {author} {\bibinfo {author} {\bibfnamefont {J.-J.}\ \bibnamefont {Chen}}, \bibinfo {author} {\bibfnamefont {J.}~\bibnamefont {Cui}}, \bibinfo {author} {\bibfnamefont {Y.-R.}\ \bibnamefont {Zhang}},\ and\ \bibinfo {author} {\bibfnamefont {H.}~\bibnamefont {Fan}},\ }\bibfield  {title} {\bibinfo {title} {Coherence susceptibility as a probe of quantum phase transitions},\ }\href@noop {} {\bibfield  {journal} {\bibinfo  {journal} {Phys. Rev. A}\ }\textbf {\bibinfo {volume} {94}},\ \bibinfo {pages} {022112} (\bibinfo {year} {2016})}\BibitemShut {NoStop}%
\bibitem [{\citenamefont {Hu}\ and\ \citenamefont {Fan}(2017)}]{hu2017relative}%
  \BibitemOpen
  \bibfield  {author} {\bibinfo {author} {\bibfnamefont {M.-L.}\ \bibnamefont {Hu}}\ and\ \bibinfo {author} {\bibfnamefont {H.}~\bibnamefont {Fan}},\ }\bibfield  {title} {\bibinfo {title} {Relative quantum coherence, incompatibility, and quantum correlations of states},\ }\href@noop {} {\bibfield  {journal} {\bibinfo  {journal} {Phys. Rev. A}\ }\textbf {\bibinfo {volume} {95}},\ \bibinfo {pages} {052106} (\bibinfo {year} {2017})}\BibitemShut {NoStop}%
\bibitem [{\citenamefont {Hu}\ \emph {et~al.}(2018)\citenamefont {Hu}, \citenamefont {Hu}, \citenamefont {Wang}, \citenamefont {Peng}, \citenamefont {Zhang},\ and\ \citenamefont {Fan}}]{hu2018quantum}%
  \BibitemOpen
  \bibfield  {author} {\bibinfo {author} {\bibfnamefont {M.-L.}\ \bibnamefont {Hu}}, \bibinfo {author} {\bibfnamefont {X.}~\bibnamefont {Hu}}, \bibinfo {author} {\bibfnamefont {J.}~\bibnamefont {Wang}}, \bibinfo {author} {\bibfnamefont {Y.}~\bibnamefont {Peng}}, \bibinfo {author} {\bibfnamefont {Y.-R.}\ \bibnamefont {Zhang}},\ and\ \bibinfo {author} {\bibfnamefont {H.}~\bibnamefont {Fan}},\ }\bibfield  {title} {\bibinfo {title} {Quantum coherence and geometric quantum discord},\ }\href@noop {} {\bibfield  {journal} {\bibinfo  {journal} {Phys. Rep.}\ }\textbf {\bibinfo {volume} {762}},\ \bibinfo {pages} {1} (\bibinfo {year} {2018})}\BibitemShut {NoStop}%
\bibitem [{\citenamefont {Chitambar}\ and\ \citenamefont {Gour}(2016)}]{chitambar2016comparison}%
  \BibitemOpen
  \bibfield  {author} {\bibinfo {author} {\bibfnamefont {E.}~\bibnamefont {Chitambar}}\ and\ \bibinfo {author} {\bibfnamefont {G.}~\bibnamefont {Gour}},\ }\bibfield  {title} {\bibinfo {title} {Comparison of incoherent operations and measures of coherence},\ }\href@noop {} {\bibfield  {journal} {\bibinfo  {journal} {Phys. Rev. A}\ }\textbf {\bibinfo {volume} {94}},\ \bibinfo {pages} {052336} (\bibinfo {year} {2016})}\BibitemShut {NoStop}%
\bibitem [{\citenamefont {Napoli}\ \emph {et~al.}(2016)\citenamefont {Napoli}, \citenamefont {Bromley}, \citenamefont {Cianciaruso}, \citenamefont {Piani}, \citenamefont {Johnston},\ and\ \citenamefont {Adesso}}]{napoli2016robustness}%
  \BibitemOpen
  \bibfield  {author} {\bibinfo {author} {\bibfnamefont {C.}~\bibnamefont {Napoli}}, \bibinfo {author} {\bibfnamefont {T.~R.}\ \bibnamefont {Bromley}}, \bibinfo {author} {\bibfnamefont {M.}~\bibnamefont {Cianciaruso}}, \bibinfo {author} {\bibfnamefont {M.}~\bibnamefont {Piani}}, \bibinfo {author} {\bibfnamefont {N.}~\bibnamefont {Johnston}},\ and\ \bibinfo {author} {\bibfnamefont {G.}~\bibnamefont {Adesso}},\ }\bibfield  {title} {\bibinfo {title} {Robustness of coherence: an operational and observable measure of quantum coherence},\ }\href@noop {} {\bibfield  {journal} {\bibinfo  {journal} {Phys. Rev. Lett.}\ }\textbf {\bibinfo {volume} {116}},\ \bibinfo {pages} {150502} (\bibinfo {year} {2016})}\BibitemShut {NoStop}%
\bibitem [{\citenamefont {Rana}\ \emph {et~al.}(2016)\citenamefont {Rana}, \citenamefont {Parashar},\ and\ \citenamefont {Lewenstein}}]{rana2016trace}%
  \BibitemOpen
  \bibfield  {author} {\bibinfo {author} {\bibfnamefont {S.}~\bibnamefont {Rana}}, \bibinfo {author} {\bibfnamefont {P.}~\bibnamefont {Parashar}},\ and\ \bibinfo {author} {\bibfnamefont {M.}~\bibnamefont {Lewenstein}},\ }\bibfield  {title} {\bibinfo {title} {Trace-distance measure of coherence},\ }\href@noop {} {\bibfield  {journal} {\bibinfo  {journal} {Phys. Rev. A}\ }\textbf {\bibinfo {volume} {93}},\ \bibinfo {pages} {012110} (\bibinfo {year} {2016})}\BibitemShut {NoStop}%
\bibitem [{\citenamefont {Yu}\ \emph {et~al.}(2016)\citenamefont {Yu}, \citenamefont {Zhang}, \citenamefont {Xu},\ and\ \citenamefont {Tong}}]{yu2016alternative}%
  \BibitemOpen
  \bibfield  {author} {\bibinfo {author} {\bibfnamefont {X.-D.}\ \bibnamefont {Yu}}, \bibinfo {author} {\bibfnamefont {D.-J.}\ \bibnamefont {Zhang}}, \bibinfo {author} {\bibfnamefont {G.}~\bibnamefont {Xu}},\ and\ \bibinfo {author} {\bibfnamefont {D.}~\bibnamefont {Tong}},\ }\bibfield  {title} {\bibinfo {title} {Alternative framework for quantifying coherence},\ }\href@noop {} {\bibfield  {journal} {\bibinfo  {journal} {Phys. Rev. A}\ }\textbf {\bibinfo {volume} {94}},\ \bibinfo {pages} {060302} (\bibinfo {year} {2016})}\BibitemShut {NoStop}%
\bibitem [{\citenamefont {Streltsov}\ \emph {et~al.}(2017)\citenamefont {Streltsov}, \citenamefont {Adesso},\ and\ \citenamefont {Plenio}}]{streltsov2017colloquium}%
  \BibitemOpen
  \bibfield  {author} {\bibinfo {author} {\bibfnamefont {A.}~\bibnamefont {Streltsov}}, \bibinfo {author} {\bibfnamefont {G.}~\bibnamefont {Adesso}},\ and\ \bibinfo {author} {\bibfnamefont {M.~B.}\ \bibnamefont {Plenio}},\ }\bibfield  {title} {\bibinfo {title} {Colloquium: Quantum coherence as a resource},\ }\href@noop {} {\bibfield  {journal} {\bibinfo  {journal} {Rev. Mod. Phys.}\ }\textbf {\bibinfo {volume} {89}},\ \bibinfo {pages} {041003} (\bibinfo {year} {2017})}\BibitemShut {NoStop}%
\bibitem [{\citenamefont {Baumgratz}\ \emph {et~al.}(2014)\citenamefont {Baumgratz}, \citenamefont {Cramer},\ and\ \citenamefont {Plenio}}]{baumgratz2014quantifying}%
  \BibitemOpen
  \bibfield  {author} {\bibinfo {author} {\bibfnamefont {T.}~\bibnamefont {Baumgratz}}, \bibinfo {author} {\bibfnamefont {M.}~\bibnamefont {Cramer}},\ and\ \bibinfo {author} {\bibfnamefont {M.~B.}\ \bibnamefont {Plenio}},\ }\bibfield  {title} {\bibinfo {title} {Quantifying coherence},\ }\href@noop {} {\bibfield  {journal} {\bibinfo  {journal} {Phys. Rev. Lett.}\ }\textbf {\bibinfo {volume} {113}},\ \bibinfo {pages} {140401} (\bibinfo {year} {2014})}\BibitemShut {NoStop}%
\bibitem [{\citenamefont {Winter}\ and\ \citenamefont {Yang}(2016)}]{winter2016operational}%
  \BibitemOpen
  \bibfield  {author} {\bibinfo {author} {\bibfnamefont {A.}~\bibnamefont {Winter}}\ and\ \bibinfo {author} {\bibfnamefont {D.}~\bibnamefont {Yang}},\ }\bibfield  {title} {\bibinfo {title} {Operational resource theory of coherence},\ }\href@noop {} {\bibfield  {journal} {\bibinfo  {journal} {Phys. Rev. Lett.}\ }\textbf {\bibinfo {volume} {116}},\ \bibinfo {pages} {120404} (\bibinfo {year} {2016})}\BibitemShut {NoStop}%
\bibitem [{\citenamefont {Chitambar}\ and\ \citenamefont {Hsieh}(2016)}]{chitambar2016relating}%
  \BibitemOpen
  \bibfield  {author} {\bibinfo {author} {\bibfnamefont {E.}~\bibnamefont {Chitambar}}\ and\ \bibinfo {author} {\bibfnamefont {M.-H.}\ \bibnamefont {Hsieh}},\ }\bibfield  {title} {\bibinfo {title} {Relating the resource theories of entanglement and quantum coherence},\ }\href@noop {} {\bibfield  {journal} {\bibinfo  {journal} {Phys. Rev. Lett.}\ }\textbf {\bibinfo {volume} {117}},\ \bibinfo {pages} {020402} (\bibinfo {year} {2016})}\BibitemShut {NoStop}%
\bibitem [{\citenamefont {Giovannetti}\ \emph {et~al.}(2004)\citenamefont {Giovannetti}, \citenamefont {Lloyd},\ and\ \citenamefont {Maccone}}]{giovannetti2004quantum}%
  \BibitemOpen
  \bibfield  {author} {\bibinfo {author} {\bibfnamefont {V.}~\bibnamefont {Giovannetti}}, \bibinfo {author} {\bibfnamefont {S.}~\bibnamefont {Lloyd}},\ and\ \bibinfo {author} {\bibfnamefont {L.}~\bibnamefont {Maccone}},\ }\bibfield  {title} {\bibinfo {title} {Quantum-enhanced measurements: beating the standard quantum limit},\ }\href@noop {} {\bibfield  {journal} {\bibinfo  {journal} {Science}\ }\textbf {\bibinfo {volume} {306}},\ \bibinfo {pages} {1330} (\bibinfo {year} {2004})}\BibitemShut {NoStop}%
\bibitem [{\citenamefont {Demkowicz-Dobrza{\'n}ski}\ and\ \citenamefont {Maccone}(2014)}]{demkowicz2014using}%
  \BibitemOpen
  \bibfield  {author} {\bibinfo {author} {\bibfnamefont {R.}~\bibnamefont {Demkowicz-Dobrza{\'n}ski}}\ and\ \bibinfo {author} {\bibfnamefont {L.}~\bibnamefont {Maccone}},\ }\bibfield  {title} {\bibinfo {title} {Using entanglement against noise in quantum metrology},\ }\href@noop {} {\bibfield  {journal} {\bibinfo  {journal} {Phys. Rev. Lett.}\ }\textbf {\bibinfo {volume} {113}},\ \bibinfo {pages} {250801} (\bibinfo {year} {2014})}\BibitemShut {NoStop}%
\bibitem [{\citenamefont {Sun}\ \emph {et~al.}(2022)\citenamefont {Sun}, \citenamefont {Liu}, \citenamefont {Wang}, \citenamefont {Hao}, \citenamefont {Xu}, \citenamefont {Xu}, \citenamefont {Li}, \citenamefont {Guo}, \citenamefont {Castellini}, \citenamefont {Lami}, \citenamefont {Winter}, \citenamefont {Adesso}, \citenamefont {Compagno},\ and\ \citenamefont {{Lo Franco}}}]{SunPNAS}%
  \BibitemOpen
  \bibfield  {author} {\bibinfo {author} {\bibfnamefont {K.}~\bibnamefont {Sun}}, \bibinfo {author} {\bibfnamefont {Z.-H.}\ \bibnamefont {Liu}}, \bibinfo {author} {\bibfnamefont {Y.}~\bibnamefont {Wang}}, \bibinfo {author} {\bibfnamefont {Z.-Y.}\ \bibnamefont {Hao}}, \bibinfo {author} {\bibfnamefont {X.-Y.}\ \bibnamefont {Xu}}, \bibinfo {author} {\bibfnamefont {J.-S.}\ \bibnamefont {Xu}}, \bibinfo {author} {\bibfnamefont {C.-F.}\ \bibnamefont {Li}}, \bibinfo {author} {\bibfnamefont {G.-C.}\ \bibnamefont {Guo}}, \bibinfo {author} {\bibfnamefont {A.}~\bibnamefont {Castellini}}, \bibinfo {author} {\bibfnamefont {L.}~\bibnamefont {Lami}}, \bibinfo {author} {\bibfnamefont {A.}~\bibnamefont {Winter}}, \bibinfo {author} {\bibfnamefont {G.}~\bibnamefont {Adesso}}, \bibinfo {author} {\bibfnamefont {G.}~\bibnamefont {Compagno}},\ and\ \bibinfo {author} {\bibfnamefont {R.}~\bibnamefont {{Lo Franco}}},\ }\bibfield  {title} {\bibinfo {title} {Activation of indistinguishability-based quantum coherence for enhanced
  metrological applications with particle statistics imprint},\ }\href@noop {} {\bibfield  {journal} {\bibinfo  {journal} {Proceedings of the National Academy of Sciences}\ }\textbf {\bibinfo {volume} {119}},\ \bibinfo {pages} {e2119765119} (\bibinfo {year} {2022})}\BibitemShut {NoStop}%
\bibitem [{\citenamefont {Asb{\'o}th}\ \emph {et~al.}(2005)\citenamefont {Asb{\'o}th}, \citenamefont {Calsamiglia},\ and\ \citenamefont {Ritsch}}]{asboth2005computable}%
  \BibitemOpen
  \bibfield  {author} {\bibinfo {author} {\bibfnamefont {J.~K.}\ \bibnamefont {Asb{\'o}th}}, \bibinfo {author} {\bibfnamefont {J.}~\bibnamefont {Calsamiglia}},\ and\ \bibinfo {author} {\bibfnamefont {H.}~\bibnamefont {Ritsch}},\ }\bibfield  {title} {\bibinfo {title} {Computable measure of nonclassicality for light},\ }\href@noop {} {\bibfield  {journal} {\bibinfo  {journal} {Phys. Rev. Lett.}\ }\textbf {\bibinfo {volume} {94}},\ \bibinfo {pages} {173602} (\bibinfo {year} {2005})}\BibitemShut {NoStop}%
\bibitem [{\citenamefont {Bennett}\ and\ \citenamefont {Wiesner}(1992)}]{bennett1992communication}%
  \BibitemOpen
  \bibfield  {author} {\bibinfo {author} {\bibfnamefont {C.~H.}\ \bibnamefont {Bennett}}\ and\ \bibinfo {author} {\bibfnamefont {S.~J.}\ \bibnamefont {Wiesner}},\ }\bibfield  {title} {\bibinfo {title} {Communication via one-and two-particle operators on einstein-podolsky-rosen states},\ }\href@noop {} {\bibfield  {journal} {\bibinfo  {journal} {Phys. Rev. Lett.}\ }\textbf {\bibinfo {volume} {69}},\ \bibinfo {pages} {2881} (\bibinfo {year} {1992})}\BibitemShut {NoStop}%
\bibitem [{\citenamefont {Laflamme}\ \emph {et~al.}(1996)\citenamefont {Laflamme}, \citenamefont {Miquel}, \citenamefont {Paz},\ and\ \citenamefont {Zurek}}]{laflamme1996perfect}%
  \BibitemOpen
  \bibfield  {author} {\bibinfo {author} {\bibfnamefont {R.}~\bibnamefont {Laflamme}}, \bibinfo {author} {\bibfnamefont {C.}~\bibnamefont {Miquel}}, \bibinfo {author} {\bibfnamefont {J.~P.}\ \bibnamefont {Paz}},\ and\ \bibinfo {author} {\bibfnamefont {W.~H.}\ \bibnamefont {Zurek}},\ }\bibfield  {title} {\bibinfo {title} {Perfect quantum error correcting code},\ }\href@noop {} {\bibfield  {journal} {\bibinfo  {journal} {Phys. Rev. Lett.}\ }\textbf {\bibinfo {volume} {77}},\ \bibinfo {pages} {198} (\bibinfo {year} {1996})}\BibitemShut {NoStop}%
\bibitem [{\citenamefont {Plenio}\ \emph {et~al.}(1997)\citenamefont {Plenio}, \citenamefont {Vedral},\ and\ \citenamefont {Knight}}]{plenio1997quantum}%
  \BibitemOpen
  \bibfield  {author} {\bibinfo {author} {\bibfnamefont {M.~B.}\ \bibnamefont {Plenio}}, \bibinfo {author} {\bibfnamefont {V.}~\bibnamefont {Vedral}},\ and\ \bibinfo {author} {\bibfnamefont {P.~L.}\ \bibnamefont {Knight}},\ }\bibfield  {title} {\bibinfo {title} {Quantum error correction in the presence of spontaneous emission},\ }\href@noop {} {\bibfield  {journal} {\bibinfo  {journal} {Phys. Rev. A}\ }\textbf {\bibinfo {volume} {55}},\ \bibinfo {pages} {67} (\bibinfo {year} {1997})}\BibitemShut {NoStop}%
\bibitem [{\citenamefont {Ekert}(1991)}]{ekert1991quantum}%
  \BibitemOpen
  \bibfield  {author} {\bibinfo {author} {\bibfnamefont {A.~K.}\ \bibnamefont {Ekert}},\ }\bibfield  {title} {\bibinfo {title} {Quantum cryptography based on bell’s theorem},\ }\href@noop {} {\bibfield  {journal} {\bibinfo  {journal} {Phys. Rev. Lett.}\ }\textbf {\bibinfo {volume} {67}},\ \bibinfo {pages} {661} (\bibinfo {year} {1991})}\BibitemShut {NoStop}%
\bibitem [{\citenamefont {Unruh}(1995)}]{unruh1995maintaining}%
  \BibitemOpen
  \bibfield  {author} {\bibinfo {author} {\bibfnamefont {W.~G.}\ \bibnamefont {Unruh}},\ }\bibfield  {title} {\bibinfo {title} {Maintaining coherence in quantum computers},\ }\href@noop {} {\bibfield  {journal} {\bibinfo  {journal} {Phys. Rev. A}\ }\textbf {\bibinfo {volume} {51}},\ \bibinfo {pages} {992} (\bibinfo {year} {1995})}\BibitemShut {NoStop}%
\bibitem [{\citenamefont {Duan}\ and\ \citenamefont {Guo}(1997)}]{duan1997preserving}%
  \BibitemOpen
  \bibfield  {author} {\bibinfo {author} {\bibfnamefont {L.-M.}\ \bibnamefont {Duan}}\ and\ \bibinfo {author} {\bibfnamefont {G.-C.}\ \bibnamefont {Guo}},\ }\bibfield  {title} {\bibinfo {title} {Preserving coherence in quantum computation by pairing quantum bits},\ }\href@noop {} {\bibfield  {journal} {\bibinfo  {journal} {Phys. Rev. Lett.}\ }\textbf {\bibinfo {volume} {79}},\ \bibinfo {pages} {1953} (\bibinfo {year} {1997})}\BibitemShut {NoStop}%
\bibitem [{\citenamefont {Wang}(2001)}]{wang2001entanglement}%
  \BibitemOpen
  \bibfield  {author} {\bibinfo {author} {\bibfnamefont {X.}~\bibnamefont {Wang}},\ }\bibfield  {title} {\bibinfo {title} {Entanglement in the quantum heisenberg xy model},\ }\href@noop {} {\bibfield  {journal} {\bibinfo  {journal} {Phys. Rev. A}\ }\textbf {\bibinfo {volume} {64}},\ \bibinfo {pages} {012313} (\bibinfo {year} {2001})}\BibitemShut {NoStop}%
\bibitem [{\citenamefont {Arnesen}\ \emph {et~al.}(2001)\citenamefont {Arnesen}, \citenamefont {Bose},\ and\ \citenamefont {Vedral}}]{arnesen2001natural}%
  \BibitemOpen
  \bibfield  {author} {\bibinfo {author} {\bibfnamefont {M.}~\bibnamefont {Arnesen}}, \bibinfo {author} {\bibfnamefont {S.}~\bibnamefont {Bose}},\ and\ \bibinfo {author} {\bibfnamefont {V.}~\bibnamefont {Vedral}},\ }\bibfield  {title} {\bibinfo {title} {Natural thermal and magnetic entanglement in the 1d heisenberg model},\ }\href@noop {} {\bibfield  {journal} {\bibinfo  {journal} {Phys. Rev. Lett.}\ }\textbf {\bibinfo {volume} {87}},\ \bibinfo {pages} {017901} (\bibinfo {year} {2001})}\BibitemShut {NoStop}%
\bibitem [{\citenamefont {Giovannetti}\ \emph {et~al.}(2011)\citenamefont {Giovannetti}, \citenamefont {Lloyd},\ and\ \citenamefont {Maccone}}]{giovannetti2011advances}%
  \BibitemOpen
  \bibfield  {author} {\bibinfo {author} {\bibfnamefont {V.}~\bibnamefont {Giovannetti}}, \bibinfo {author} {\bibfnamefont {S.}~\bibnamefont {Lloyd}},\ and\ \bibinfo {author} {\bibfnamefont {L.}~\bibnamefont {Maccone}},\ }\bibfield  {title} {\bibinfo {title} {Advances in quantum metrology},\ }\href@noop {} {\bibfield  {journal} {\bibinfo  {journal} {Nat. Photonics}\ }\textbf {\bibinfo {volume} {5}},\ \bibinfo {pages} {222} (\bibinfo {year} {2011})}\BibitemShut {NoStop}%
\bibitem [{\citenamefont {Joo}\ \emph {et~al.}(2011)\citenamefont {Joo}, \citenamefont {Munro},\ and\ \citenamefont {Spiller}}]{joo2011quantum}%
  \BibitemOpen
  \bibfield  {author} {\bibinfo {author} {\bibfnamefont {J.}~\bibnamefont {Joo}}, \bibinfo {author} {\bibfnamefont {W.~J.}\ \bibnamefont {Munro}},\ and\ \bibinfo {author} {\bibfnamefont {T.~P.}\ \bibnamefont {Spiller}},\ }\bibfield  {title} {\bibinfo {title} {Quantum metrology with entangled coherent states},\ }\href@noop {} {\bibfield  {journal} {\bibinfo  {journal} {Phys. Rev. Lett.}\ }\textbf {\bibinfo {volume} {107}},\ \bibinfo {pages} {083601} (\bibinfo {year} {2011})}\BibitemShut {NoStop}%
\bibitem [{\citenamefont {Liu}\ \emph {et~al.}(2013)\citenamefont {Liu}, \citenamefont {Jing},\ and\ \citenamefont {Wang}}]{liu2013phase}%
  \BibitemOpen
  \bibfield  {author} {\bibinfo {author} {\bibfnamefont {J.}~\bibnamefont {Liu}}, \bibinfo {author} {\bibfnamefont {X.}~\bibnamefont {Jing}},\ and\ \bibinfo {author} {\bibfnamefont {X.}~\bibnamefont {Wang}},\ }\bibfield  {title} {\bibinfo {title} {Phase-matching condition for enhancement of phase sensitivity in quantum metrology},\ }\href@noop {} {\bibfield  {journal} {\bibinfo  {journal} {Phys. Rev. A}\ }\textbf {\bibinfo {volume} {88}},\ \bibinfo {pages} {042316} (\bibinfo {year} {2013})}\BibitemShut {NoStop}%
\bibitem [{\citenamefont {Chaves}\ \emph {et~al.}(2013)\citenamefont {Chaves}, \citenamefont {Brask}, \citenamefont {Markiewicz}, \citenamefont {Ko{\l}ody{\'n}ski},\ and\ \citenamefont {Ac{\'\i}n}}]{chaves2013noisy}%
  \BibitemOpen
  \bibfield  {author} {\bibinfo {author} {\bibfnamefont {R.}~\bibnamefont {Chaves}}, \bibinfo {author} {\bibfnamefont {J.}~\bibnamefont {Brask}}, \bibinfo {author} {\bibfnamefont {M.}~\bibnamefont {Markiewicz}}, \bibinfo {author} {\bibfnamefont {J.}~\bibnamefont {Ko{\l}ody{\'n}ski}},\ and\ \bibinfo {author} {\bibfnamefont {A.}~\bibnamefont {Ac{\'\i}n}},\ }\bibfield  {title} {\bibinfo {title} {Noisy metrology beyond the standard quantum limit},\ }\href@noop {} {\bibfield  {journal} {\bibinfo  {journal} {Phys. Rev. Lett.}\ }\textbf {\bibinfo {volume} {111}},\ \bibinfo {pages} {120401} (\bibinfo {year} {2013})}\BibitemShut {NoStop}%
\bibitem [{\citenamefont {Zhang}\ \emph {et~al.}(2013)\citenamefont {Zhang}, \citenamefont {Li}, \citenamefont {Yang},\ and\ \citenamefont {Jin}}]{zhang2013quantum}%
  \BibitemOpen
  \bibfield  {author} {\bibinfo {author} {\bibfnamefont {Y.}~\bibnamefont {Zhang}}, \bibinfo {author} {\bibfnamefont {X.}~\bibnamefont {Li}}, \bibinfo {author} {\bibfnamefont {W.}~\bibnamefont {Yang}},\ and\ \bibinfo {author} {\bibfnamefont {G.}~\bibnamefont {Jin}},\ }\bibfield  {title} {\bibinfo {title} {Quantum fisher information of entangled coherent states in the presence of photon loss},\ }\href@noop {} {\bibfield  {journal} {\bibinfo  {journal} {Phys. Rev. A}\ }\textbf {\bibinfo {volume} {88}},\ \bibinfo {pages} {043832} (\bibinfo {year} {2013})}\BibitemShut {NoStop}%
\bibitem [{\citenamefont {Alipour}\ \emph {et~al.}(2014)\citenamefont {Alipour}, \citenamefont {Mehboudi},\ and\ \citenamefont {Rezakhani}}]{alipour2014quantum}%
  \BibitemOpen
  \bibfield  {author} {\bibinfo {author} {\bibfnamefont {S.}~\bibnamefont {Alipour}}, \bibinfo {author} {\bibfnamefont {M.}~\bibnamefont {Mehboudi}},\ and\ \bibinfo {author} {\bibfnamefont {A.}~\bibnamefont {Rezakhani}},\ }\bibfield  {title} {\bibinfo {title} {Quantum metrology in open systems: dissipative cram{\'e}r-rao bound},\ }\href@noop {} {\bibfield  {journal} {\bibinfo  {journal} {Phys. Rev. Lett.}\ }\textbf {\bibinfo {volume} {112}},\ \bibinfo {pages} {120405} (\bibinfo {year} {2014})}\BibitemShut {NoStop}%
\bibitem [{\citenamefont {Lu}\ \emph {et~al.}(2015)\citenamefont {Lu}, \citenamefont {Yu},\ and\ \citenamefont {Oh}}]{lu2015robust}%
  \BibitemOpen
  \bibfield  {author} {\bibinfo {author} {\bibfnamefont {X.-M.}\ \bibnamefont {Lu}}, \bibinfo {author} {\bibfnamefont {S.}~\bibnamefont {Yu}},\ and\ \bibinfo {author} {\bibfnamefont {C.}~\bibnamefont {Oh}},\ }\bibfield  {title} {\bibinfo {title} {Robust quantum metrological schemes based on protection of quantum fisher information},\ }\href@noop {} {\bibfield  {journal} {\bibinfo  {journal} {Nat. Commun.}\ }\textbf {\bibinfo {volume} {6}},\ \bibinfo {pages} {7282} (\bibinfo {year} {2015})}\BibitemShut {NoStop}%
\bibitem [{\citenamefont {Correa}\ \emph {et~al.}(2015)\citenamefont {Correa}, \citenamefont {Mehboudi}, \citenamefont {Adesso},\ and\ \citenamefont {Sanpera}}]{correa2015individual}%
  \BibitemOpen
  \bibfield  {author} {\bibinfo {author} {\bibfnamefont {L.~A.}\ \bibnamefont {Correa}}, \bibinfo {author} {\bibfnamefont {M.}~\bibnamefont {Mehboudi}}, \bibinfo {author} {\bibfnamefont {G.}~\bibnamefont {Adesso}},\ and\ \bibinfo {author} {\bibfnamefont {A.}~\bibnamefont {Sanpera}},\ }\bibfield  {title} {\bibinfo {title} {Individual quantum probes for optimal thermometry},\ }\href@noop {} {\bibfield  {journal} {\bibinfo  {journal} {Phys. Rev. Lett.}\ }\textbf {\bibinfo {volume} {114}},\ \bibinfo {pages} {220405} (\bibinfo {year} {2015})}\BibitemShut {NoStop}%
\bibitem [{\citenamefont {Baumgratz}\ and\ \citenamefont {Datta}(2016)}]{baumgratz2016quantum}%
  \BibitemOpen
  \bibfield  {author} {\bibinfo {author} {\bibfnamefont {T.}~\bibnamefont {Baumgratz}}\ and\ \bibinfo {author} {\bibfnamefont {A.}~\bibnamefont {Datta}},\ }\bibfield  {title} {\bibinfo {title} {Quantum enhanced estimation of a multidimensional field},\ }\href@noop {} {\bibfield  {journal} {\bibinfo  {journal} {Phys. Rev. Lett.}\ }\textbf {\bibinfo {volume} {116}},\ \bibinfo {pages} {030801} (\bibinfo {year} {2016})}\BibitemShut {NoStop}%
\bibitem [{\citenamefont {Liu}\ and\ \citenamefont {Yuan}(2017)}]{liu2017quantum}%
  \BibitemOpen
  \bibfield  {author} {\bibinfo {author} {\bibfnamefont {J.}~\bibnamefont {Liu}}\ and\ \bibinfo {author} {\bibfnamefont {H.}~\bibnamefont {Yuan}},\ }\bibfield  {title} {\bibinfo {title} {Quantum parameter estimation with optimal control},\ }\href@noop {} {\bibfield  {journal} {\bibinfo  {journal} {Phys. Rev. A}\ }\textbf {\bibinfo {volume} {96}},\ \bibinfo {pages} {012117} (\bibinfo {year} {2017})}\BibitemShut {NoStop}%
\bibitem [{\citenamefont {Liu}\ and\ \citenamefont {Fan}(2023)}]{liu2023application}%
  \BibitemOpen
  \bibfield  {author} {\bibinfo {author} {\bibfnamefont {S.-Y.}\ \bibnamefont {Liu}}\ and\ \bibinfo {author} {\bibfnamefont {H.}~\bibnamefont {Fan}},\ }\bibfield  {title} {\bibinfo {title} {The application of quantum coherence as a resource},\ }\href@noop {} {\bibfield  {journal} {\bibinfo  {journal} {Chin. Phys. B}\ }\textbf {\bibinfo {volume} {32}},\ \bibinfo {pages} {110304} (\bibinfo {year} {2023})}\BibitemShut {NoStop}%
\bibitem [{\citenamefont {Shahandeh}\ \emph {et~al.}(2019)\citenamefont {Shahandeh}, \citenamefont {Lund},\ and\ \citenamefont {Ralph}}]{shahandeh2019quantum}%
  \BibitemOpen
  \bibfield  {author} {\bibinfo {author} {\bibfnamefont {F.}~\bibnamefont {Shahandeh}}, \bibinfo {author} {\bibfnamefont {A.~P.}\ \bibnamefont {Lund}},\ and\ \bibinfo {author} {\bibfnamefont {T.~C.}\ \bibnamefont {Ralph}},\ }\bibfield  {title} {\bibinfo {title} {Quantum correlations and global coherence in distributed quantum computing},\ }\href@noop {} {\bibfield  {journal} {\bibinfo  {journal} {Phys. Rev. A}\ }\textbf {\bibinfo {volume} {99}},\ \bibinfo {pages} {052303} (\bibinfo {year} {2019})}\BibitemShut {NoStop}%
\bibitem [{\citenamefont {Breuer}\ \emph {et~al.}(2002)\citenamefont {Breuer}, \citenamefont {Petruccione} \emph {et~al.}}]{breuer2002theory}%
  \BibitemOpen
  \bibfield  {author} {\bibinfo {author} {\bibfnamefont {H.-P.}\ \bibnamefont {Breuer}}, \bibinfo {author} {\bibfnamefont {F.}~\bibnamefont {Petruccione}}, \emph {et~al.},\ }\href@noop {} {\emph {\bibinfo {title} {The theory of open quantum systems}}}\ (\bibinfo  {publisher} {Oxford University Press on Demand},\ \bibinfo {year} {2002})\BibitemShut {NoStop}%
\bibitem [{\citenamefont {Zurek}(1991)}]{zurek1991decoherence}%
  \BibitemOpen
  \bibfield  {author} {\bibinfo {author} {\bibfnamefont {W.~H.}\ \bibnamefont {Zurek}},\ }\bibfield  {title} {\bibinfo {title} {Decoherence and the transition from quantum to classical},\ }\href@noop {} {\bibfield  {journal} {\bibinfo  {journal} {Phys. Today}\ }\textbf {\bibinfo {volume} {44}},\ \bibinfo {pages} {36} (\bibinfo {year} {1991})}\BibitemShut {NoStop}%
\bibitem [{\citenamefont {Zurek}\ and\ \citenamefont {Paz}(1994)}]{zurek1994decoherence}%
  \BibitemOpen
  \bibfield  {author} {\bibinfo {author} {\bibfnamefont {W.~H.}\ \bibnamefont {Zurek}}\ and\ \bibinfo {author} {\bibfnamefont {J.~P.}\ \bibnamefont {Paz}},\ }\bibfield  {title} {\bibinfo {title} {Decoherence, chaos, and the second law},\ }\href@noop {} {\bibfield  {journal} {\bibinfo  {journal} {Phys. Rev. Lett.}\ }\textbf {\bibinfo {volume} {72}},\ \bibinfo {pages} {2508} (\bibinfo {year} {1994})}\BibitemShut {NoStop}%
\bibitem [{\citenamefont {{Lo Franco}}\ \emph {et~al.}(2013)\citenamefont {{Lo Franco}}, \citenamefont {Bellomo}, \citenamefont {Maniscalco},\ and\ \citenamefont {Compagno}}]{lofranco2013dynamics}%
  \BibitemOpen
  \bibfield  {author} {\bibinfo {author} {\bibfnamefont {R.}~\bibnamefont {{Lo Franco}}}, \bibinfo {author} {\bibfnamefont {B.}~\bibnamefont {Bellomo}}, \bibinfo {author} {\bibfnamefont {S.}~\bibnamefont {Maniscalco}},\ and\ \bibinfo {author} {\bibfnamefont {G.}~\bibnamefont {Compagno}},\ }\bibfield  {title} {\bibinfo {title} {Dynamics of quantum correlations in two-qubit systems within non-markovian environments},\ }\href@noop {} {\bibfield  {journal} {\bibinfo  {journal} {Int. J. Mod. Phys. B}\ }\textbf {\bibinfo {volume} {27}},\ \bibinfo {pages} {1345053} (\bibinfo {year} {2013})}\BibitemShut {NoStop}%
\bibitem [{\citenamefont {Mortezapour}\ \emph {et~al.}(2018)\citenamefont {Mortezapour}, \citenamefont {Naeimi},\ and\ \citenamefont {Lo~Franco}}]{mortezapour2018coherence}%
  \BibitemOpen
  \bibfield  {author} {\bibinfo {author} {\bibfnamefont {A.}~\bibnamefont {Mortezapour}}, \bibinfo {author} {\bibfnamefont {G.}~\bibnamefont {Naeimi}},\ and\ \bibinfo {author} {\bibfnamefont {R.}~\bibnamefont {Lo~Franco}},\ }\bibfield  {title} {\bibinfo {title} {Coherence and entanglement dynamics of vibrating qubits},\ }\href@noop {} {\bibfield  {journal} {\bibinfo  {journal} {Opt. Commun.}\ }\textbf {\bibinfo {volume} {424}},\ \bibinfo {pages} {26} (\bibinfo {year} {2018})}\BibitemShut {NoStop}%
\bibitem [{\citenamefont {Scala}\ \emph {et~al.}(2008)\citenamefont {Scala}, \citenamefont {Militello}, \citenamefont {Messina}, \citenamefont {Maniscalco}, \citenamefont {Piilo},\ and\ \citenamefont {Suominen}}]{scala2008population}%
  \BibitemOpen
  \bibfield  {author} {\bibinfo {author} {\bibfnamefont {M.}~\bibnamefont {Scala}}, \bibinfo {author} {\bibfnamefont {B.}~\bibnamefont {Militello}}, \bibinfo {author} {\bibfnamefont {A.}~\bibnamefont {Messina}}, \bibinfo {author} {\bibfnamefont {S.}~\bibnamefont {Maniscalco}}, \bibinfo {author} {\bibfnamefont {J.}~\bibnamefont {Piilo}},\ and\ \bibinfo {author} {\bibfnamefont {K.-A.}\ \bibnamefont {Suominen}},\ }\bibfield  {title} {\bibinfo {title} {Population trapping due to cavity losses},\ }\href@noop {} {\bibfield  {journal} {\bibinfo  {journal} {Phys. Rev. A}\ }\textbf {\bibinfo {volume} {77}},\ \bibinfo {pages} {043827} (\bibinfo {year} {2008})}\BibitemShut {NoStop}%
\bibitem [{\citenamefont {Viola}\ and\ \citenamefont {Lloyd}(1998)}]{viola1998dynamical}%
  \BibitemOpen
  \bibfield  {author} {\bibinfo {author} {\bibfnamefont {L.}~\bibnamefont {Viola}}\ and\ \bibinfo {author} {\bibfnamefont {S.}~\bibnamefont {Lloyd}},\ }\bibfield  {title} {\bibinfo {title} {Dynamical suppression of decoherence in two-state quantum systems},\ }\href@noop {} {\bibfield  {journal} {\bibinfo  {journal} {Phys. Rev. A}\ }\textbf {\bibinfo {volume} {58}},\ \bibinfo {pages} {2733} (\bibinfo {year} {1998})}\BibitemShut {NoStop}%
\bibitem [{\citenamefont {Branderhorst}\ \emph {et~al.}(2008)\citenamefont {Branderhorst}, \citenamefont {Londero}, \citenamefont {Wasylczyk}, \citenamefont {Brif}, \citenamefont {Kosut}, \citenamefont {Rabitz},\ and\ \citenamefont {Walmsley}}]{branderhorst2008coherent}%
  \BibitemOpen
  \bibfield  {author} {\bibinfo {author} {\bibfnamefont {M.~P.}\ \bibnamefont {Branderhorst}}, \bibinfo {author} {\bibfnamefont {P.}~\bibnamefont {Londero}}, \bibinfo {author} {\bibfnamefont {P.}~\bibnamefont {Wasylczyk}}, \bibinfo {author} {\bibfnamefont {C.}~\bibnamefont {Brif}}, \bibinfo {author} {\bibfnamefont {R.~L.}\ \bibnamefont {Kosut}}, \bibinfo {author} {\bibfnamefont {H.}~\bibnamefont {Rabitz}},\ and\ \bibinfo {author} {\bibfnamefont {I.~A.}\ \bibnamefont {Walmsley}},\ }\bibfield  {title} {\bibinfo {title} {Coherent control of decoherence},\ }\href@noop {} {\bibfield  {journal} {\bibinfo  {journal} {Science}\ }\textbf {\bibinfo {volume} {320}},\ \bibinfo {pages} {638} (\bibinfo {year} {2008})}\BibitemShut {NoStop}%
\bibitem [{\citenamefont {Lo~Franco}\ \emph {et~al.}(2013)\citenamefont {Lo~Franco}, \citenamefont {D'Arrigo}, \citenamefont {Falci}, \citenamefont {Compagno},\ and\ \citenamefont {Paladino}}]{franco2013spin}%
  \BibitemOpen
  \bibfield  {author} {\bibinfo {author} {\bibfnamefont {R.}~\bibnamefont {Lo~Franco}}, \bibinfo {author} {\bibfnamefont {A.}~\bibnamefont {D'Arrigo}}, \bibinfo {author} {\bibfnamefont {G.}~\bibnamefont {Falci}}, \bibinfo {author} {\bibfnamefont {G.}~\bibnamefont {Compagno}},\ and\ \bibinfo {author} {\bibfnamefont {E.}~\bibnamefont {Paladino}},\ }\bibfield  {title} {\bibinfo {title} {Spin-echo entanglement protection from random telegraph noise},\ }\href@noop {} {\bibfield  {journal} {\bibinfo  {journal} {Phys. Scr.}\ }\textbf {\bibinfo {volume} {2013}},\ \bibinfo {pages} {014043} (\bibinfo {year} {2013})}\BibitemShut {NoStop}%
\bibitem [{\citenamefont {Tan}\ \emph {et~al.}(2010)\citenamefont {Tan}, \citenamefont {Kyaw},\ and\ \citenamefont {Yeo}}]{tan2010non}%
  \BibitemOpen
  \bibfield  {author} {\bibinfo {author} {\bibfnamefont {J.}~\bibnamefont {Tan}}, \bibinfo {author} {\bibfnamefont {T.~H.}\ \bibnamefont {Kyaw}},\ and\ \bibinfo {author} {\bibfnamefont {Y.}~\bibnamefont {Yeo}},\ }\bibfield  {title} {\bibinfo {title} {Non-markovian environments and entanglement preservation},\ }\href@noop {} {\bibfield  {journal} {\bibinfo  {journal} {Phys. Rev. A}\ }\textbf {\bibinfo {volume} {81}},\ \bibinfo {pages} {062119} (\bibinfo {year} {2010})}\BibitemShut {NoStop}%
\bibitem [{\citenamefont {Scala}\ \emph {et~al.}(2011)\citenamefont {Scala}, \citenamefont {Migliore}, \citenamefont {Messina},\ and\ \citenamefont {S{\'a}nchez-Soto}}]{scala2011robust}%
  \BibitemOpen
  \bibfield  {author} {\bibinfo {author} {\bibfnamefont {M.}~\bibnamefont {Scala}}, \bibinfo {author} {\bibfnamefont {R.}~\bibnamefont {Migliore}}, \bibinfo {author} {\bibfnamefont {A.}~\bibnamefont {Messina}},\ and\ \bibinfo {author} {\bibfnamefont {L.}~\bibnamefont {S{\'a}nchez-Soto}},\ }\bibfield  {title} {\bibinfo {title} {Robust stationary entanglement of two coupled qubits in independent environments},\ }\href@noop {} {\bibfield  {journal} {\bibinfo  {journal} {Eur. Phys. J. D}\ }\textbf {\bibinfo {volume} {61}},\ \bibinfo {pages} {199} (\bibinfo {year} {2011})}\BibitemShut {NoStop}%
\bibitem [{\citenamefont {Xue}\ \emph {et~al.}(2012)\citenamefont {Xue}, \citenamefont {Wu}, \citenamefont {Zhang}, \citenamefont {Zhang}, \citenamefont {Li},\ and\ \citenamefont {Tarn}}]{xue2012decoherence}%
  \BibitemOpen
  \bibfield  {author} {\bibinfo {author} {\bibfnamefont {S.-B.}\ \bibnamefont {Xue}}, \bibinfo {author} {\bibfnamefont {R.-B.}\ \bibnamefont {Wu}}, \bibinfo {author} {\bibfnamefont {W.-M.}\ \bibnamefont {Zhang}}, \bibinfo {author} {\bibfnamefont {J.}~\bibnamefont {Zhang}}, \bibinfo {author} {\bibfnamefont {C.-W.}\ \bibnamefont {Li}},\ and\ \bibinfo {author} {\bibfnamefont {T.-J.}\ \bibnamefont {Tarn}},\ }\bibfield  {title} {\bibinfo {title} {Decoherence suppression via non-markovian coherent feedback control},\ }\href@noop {} {\bibfield  {journal} {\bibinfo  {journal} {Phys. Rev. A}\ }\textbf {\bibinfo {volume} {86}},\ \bibinfo {pages} {052304} (\bibinfo {year} {2012})}\BibitemShut {NoStop}%
\bibitem [{\citenamefont {D’Arrigo}\ \emph {et~al.}(2014)\citenamefont {D’Arrigo}, \citenamefont {Lo~Franco}, \citenamefont {Benenti}, \citenamefont {Paladino},\ and\ \citenamefont {Falci}}]{d2014recovering}%
  \BibitemOpen
  \bibfield  {author} {\bibinfo {author} {\bibfnamefont {A.}~\bibnamefont {D’Arrigo}}, \bibinfo {author} {\bibfnamefont {R.}~\bibnamefont {Lo~Franco}}, \bibinfo {author} {\bibfnamefont {G.}~\bibnamefont {Benenti}}, \bibinfo {author} {\bibfnamefont {E.}~\bibnamefont {Paladino}},\ and\ \bibinfo {author} {\bibfnamefont {G.}~\bibnamefont {Falci}},\ }\bibfield  {title} {\bibinfo {title} {Recovering entanglement by local operations},\ }\href@noop {} {\bibfield  {journal} {\bibinfo  {journal} {Ann. Phys.}\ }\textbf {\bibinfo {volume} {350}},\ \bibinfo {pages} {211} (\bibinfo {year} {2014})}\BibitemShut {NoStop}%
\bibitem [{\citenamefont {Man}\ \emph {et~al.}(2012)\citenamefont {Man}, \citenamefont {Xia},\ and\ \citenamefont {An}}]{man2012enhancing}%
  \BibitemOpen
  \bibfield  {author} {\bibinfo {author} {\bibfnamefont {Z.-X.}\ \bibnamefont {Man}}, \bibinfo {author} {\bibfnamefont {Y.-J.}\ \bibnamefont {Xia}},\ and\ \bibinfo {author} {\bibfnamefont {N.~B.}\ \bibnamefont {An}},\ }\bibfield  {title} {\bibinfo {title} {Enhancing entanglement of two qubits undergoing independent decoherences by local pre-and postmeasurements},\ }\href@noop {} {\bibfield  {journal} {\bibinfo  {journal} {Phys. Rev. A}\ }\textbf {\bibinfo {volume} {86}},\ \bibinfo {pages} {052322} (\bibinfo {year} {2012})}\BibitemShut {NoStop}%
\bibitem [{\citenamefont {Noel}\ \emph {et~al.}(1998)\citenamefont {Noel}, \citenamefont {Griffith},\ and\ \citenamefont {Gallagher}}]{noel1998frequency}%
  \BibitemOpen
  \bibfield  {author} {\bibinfo {author} {\bibfnamefont {M.~W.}\ \bibnamefont {Noel}}, \bibinfo {author} {\bibfnamefont {W.}~\bibnamefont {Griffith}},\ and\ \bibinfo {author} {\bibfnamefont {T.}~\bibnamefont {Gallagher}},\ }\bibfield  {title} {\bibinfo {title} {Frequency-modulated excitation of a two-level atom},\ }\href@noop {} {\bibfield  {journal} {\bibinfo  {journal} {Phys. Rev. A}\ }\textbf {\bibinfo {volume} {58}},\ \bibinfo {pages} {2265} (\bibinfo {year} {1998})}\BibitemShut {NoStop}%
\bibitem [{\citenamefont {Zhang}\ \emph {et~al.}(2003)\citenamefont {Zhang}, \citenamefont {Jiang}, \citenamefont {Rao},\ and\ \citenamefont {Li}}]{zhang2003frequency}%
  \BibitemOpen
  \bibfield  {author} {\bibinfo {author} {\bibfnamefont {X.}~\bibnamefont {Zhang}}, \bibinfo {author} {\bibfnamefont {H.}~\bibnamefont {Jiang}}, \bibinfo {author} {\bibfnamefont {J.}~\bibnamefont {Rao}},\ and\ \bibinfo {author} {\bibfnamefont {B.}~\bibnamefont {Li}},\ }\bibfield  {title} {\bibinfo {title} {Frequency-modulated excitation of potassium atoms},\ }\href@noop {} {\bibfield  {journal} {\bibinfo  {journal} {Phys. Rev. A}\ }\textbf {\bibinfo {volume} {68}},\ \bibinfo {pages} {025401} (\bibinfo {year} {2003})}\BibitemShut {NoStop}%
\bibitem [{\citenamefont {Silveri}\ \emph {et~al.}(2017)\citenamefont {Silveri}, \citenamefont {Tuorila}, \citenamefont {Thuneberg},\ and\ \citenamefont {Paraoanu}}]{silveri2017quantum}%
  \BibitemOpen
  \bibfield  {author} {\bibinfo {author} {\bibfnamefont {M.}~\bibnamefont {Silveri}}, \bibinfo {author} {\bibfnamefont {J.}~\bibnamefont {Tuorila}}, \bibinfo {author} {\bibfnamefont {E.}~\bibnamefont {Thuneberg}},\ and\ \bibinfo {author} {\bibfnamefont {G.}~\bibnamefont {Paraoanu}},\ }\bibfield  {title} {\bibinfo {title} {Quantum systems under frequency modulation},\ }\href@noop {} {\bibfield  {journal} {\bibinfo  {journal} {Rep. Prog. Phys.}\ }\textbf {\bibinfo {volume} {80}},\ \bibinfo {pages} {056002} (\bibinfo {year} {2017})}\BibitemShut {NoStop}%
\bibitem [{\citenamefont {Alsing}\ \emph {et~al.}(1992)\citenamefont {Alsing}, \citenamefont {Guo},\ and\ \citenamefont {Carmichael}}]{alsing1992dynamic}%
  \BibitemOpen
  \bibfield  {author} {\bibinfo {author} {\bibfnamefont {P.}~\bibnamefont {Alsing}}, \bibinfo {author} {\bibfnamefont {D.-S.}\ \bibnamefont {Guo}},\ and\ \bibinfo {author} {\bibfnamefont {H.}~\bibnamefont {Carmichael}},\ }\bibfield  {title} {\bibinfo {title} {Dynamic stark effect for the jaynes-cummings system},\ }\href@noop {} {\bibfield  {journal} {\bibinfo  {journal} {Phys. Rev. A}\ }\textbf {\bibinfo {volume} {45}},\ \bibinfo {pages} {5135} (\bibinfo {year} {1992})}\BibitemShut {NoStop}%
\bibitem [{\citenamefont {Shevchenko}\ \emph {et~al.}(2010)\citenamefont {Shevchenko}, \citenamefont {Ashhab},\ and\ \citenamefont {Nori}}]{shevchenko2010landau}%
  \BibitemOpen
  \bibfield  {author} {\bibinfo {author} {\bibfnamefont {S.~N.}\ \bibnamefont {Shevchenko}}, \bibinfo {author} {\bibfnamefont {S.}~\bibnamefont {Ashhab}},\ and\ \bibinfo {author} {\bibfnamefont {F.}~\bibnamefont {Nori}},\ }\bibfield  {title} {\bibinfo {title} {Landau--zener--st{\"u}ckelberg interferometry},\ }\href@noop {} {\bibfield  {journal} {\bibinfo  {journal} {Phys. Rep.}\ }\textbf {\bibinfo {volume} {492}},\ \bibinfo {pages} {1} (\bibinfo {year} {2010})}\BibitemShut {NoStop}%
\bibitem [{\citenamefont {Poggi}\ \emph {et~al.}(2017)\citenamefont {Poggi}, \citenamefont {Lombardo},\ and\ \citenamefont {Wisniacki}}]{poggi2017driving}%
  \BibitemOpen
  \bibfield  {author} {\bibinfo {author} {\bibfnamefont {P.~M.}\ \bibnamefont {Poggi}}, \bibinfo {author} {\bibfnamefont {F.~C.}\ \bibnamefont {Lombardo}},\ and\ \bibinfo {author} {\bibfnamefont {D.~A.}\ \bibnamefont {Wisniacki}},\ }\bibfield  {title} {\bibinfo {title} {Driving-induced amplification of non-markovianity in open quantum systems evolution},\ }\href@noop {} {\bibfield  {journal} {\bibinfo  {journal} {EPL}\ }\textbf {\bibinfo {volume} {118}},\ \bibinfo {pages} {20005} (\bibinfo {year} {2017})}\BibitemShut {NoStop}%
\bibitem [{\citenamefont {Martin}\ \emph {et~al.}(2017)\citenamefont {Martin}, \citenamefont {Refael},\ and\ \citenamefont {Halperin}}]{martin2017topological}%
  \BibitemOpen
  \bibfield  {author} {\bibinfo {author} {\bibfnamefont {I.}~\bibnamefont {Martin}}, \bibinfo {author} {\bibfnamefont {G.}~\bibnamefont {Refael}},\ and\ \bibinfo {author} {\bibfnamefont {B.}~\bibnamefont {Halperin}},\ }\bibfield  {title} {\bibinfo {title} {Topological frequency conversion in strongly driven quantum systems},\ }\href@noop {} {\bibfield  {journal} {\bibinfo  {journal} {Phys. Rev. X}\ }\textbf {\bibinfo {volume} {7}},\ \bibinfo {pages} {041008} (\bibinfo {year} {2017})}\BibitemShut {NoStop}%
\bibitem [{\citenamefont {Gray}\ \emph {et~al.}(1978)\citenamefont {Gray}, \citenamefont {Whitley},\ and\ \citenamefont {Stroud}}]{gray1978coherent}%
  \BibitemOpen
  \bibfield  {author} {\bibinfo {author} {\bibfnamefont {H.}~\bibnamefont {Gray}}, \bibinfo {author} {\bibfnamefont {R.}~\bibnamefont {Whitley}},\ and\ \bibinfo {author} {\bibfnamefont {C.}~\bibnamefont {Stroud}},\ }\bibfield  {title} {\bibinfo {title} {Coherent trapping of atomic populations},\ }\href@noop {} {\bibfield  {journal} {\bibinfo  {journal} {Optics letters}\ }\textbf {\bibinfo {volume} {3}},\ \bibinfo {pages} {218} (\bibinfo {year} {1978})}\BibitemShut {NoStop}%
\bibitem [{\citenamefont {Carrion}\ \emph {et~al.}(2024)\citenamefont {Carrion}, \citenamefont {Rojas}, \citenamefont {Filgueiras},\ and\ \citenamefont {Rojas}}]{carrion2024decoherence}%
  \BibitemOpen
  \bibfield  {author} {\bibinfo {author} {\bibfnamefont {H.~L.}\ \bibnamefont {Carrion}}, \bibinfo {author} {\bibfnamefont {O.}~\bibnamefont {Rojas}}, \bibinfo {author} {\bibfnamefont {C.}~\bibnamefont {Filgueiras}},\ and\ \bibinfo {author} {\bibfnamefont {M.}~\bibnamefont {Rojas}},\ }\bibfield  {title} {\bibinfo {title} {Decoherence effects on local quantum fisher information and quantum coherence in a spin-1/2 ising-xyz chain},\ }\href@noop {} {\bibfield  {journal} {\bibinfo  {journal} {arXiv preprint arXiv:2406.10142}\ } (\bibinfo {year} {2024})}\BibitemShut {NoStop}%
\bibitem [{\citenamefont {Huang}\ and\ \citenamefont {Situ}(2017)}]{huang2017optimal}%
  \BibitemOpen
  \bibfield  {author} {\bibinfo {author} {\bibfnamefont {Z.}~\bibnamefont {Huang}}\ and\ \bibinfo {author} {\bibfnamefont {H.}~\bibnamefont {Situ}},\ }\bibfield  {title} {\bibinfo {title} {Optimal protection of quantum coherence in noisy environment},\ }\href@noop {} {\bibfield  {journal} {\bibinfo  {journal} {Int. J. Theor. Phys.}\ }\textbf {\bibinfo {volume} {56}},\ \bibinfo {pages} {503} (\bibinfo {year} {2017})}\BibitemShut {NoStop}%
\bibitem [{\citenamefont {Wellard}\ and\ \citenamefont {Hollenberg}(2002)}]{wellard2002thermal}%
  \BibitemOpen
  \bibfield  {author} {\bibinfo {author} {\bibfnamefont {C.}~\bibnamefont {Wellard}}\ and\ \bibinfo {author} {\bibfnamefont {L.}~\bibnamefont {Hollenberg}},\ }\bibfield  {title} {\bibinfo {title} {Thermal noise in a solid state quantum computer},\ }\href@noop {} {\bibfield  {journal} {\bibinfo  {journal} {J. Phys. D}\ }\textbf {\bibinfo {volume} {35}},\ \bibinfo {pages} {2499} (\bibinfo {year} {2002})}\BibitemShut {NoStop}%
\bibitem [{\citenamefont {Simbierowicz}\ \emph {et~al.}(2024)\citenamefont {Simbierowicz}, \citenamefont {Borrelli}, \citenamefont {Monarkha}, \citenamefont {Nuutinen},\ and\ \citenamefont {Lake}}]{simbierowicz2024inherent}%
  \BibitemOpen
  \bibfield  {author} {\bibinfo {author} {\bibfnamefont {S.}~\bibnamefont {Simbierowicz}}, \bibinfo {author} {\bibfnamefont {M.}~\bibnamefont {Borrelli}}, \bibinfo {author} {\bibfnamefont {V.}~\bibnamefont {Monarkha}}, \bibinfo {author} {\bibfnamefont {V.}~\bibnamefont {Nuutinen}},\ and\ \bibinfo {author} {\bibfnamefont {R.~E.}\ \bibnamefont {Lake}},\ }\bibfield  {title} {\bibinfo {title} {Inherent thermal-noise problem in addressing qubits},\ }\href@noop {} {\bibfield  {journal} {\bibinfo  {journal} {PRX Quantum}\ }\textbf {\bibinfo {volume} {5}},\ \bibinfo {pages} {030302} (\bibinfo {year} {2024})}\BibitemShut {NoStop}%
\bibitem [{\citenamefont {Kim}\ \emph {et~al.}(2022)\citenamefont {Kim}, \citenamefont {Murugan}, \citenamefont {Olle},\ and\ \citenamefont {Rosa}}]{kim2022operator}%
  \BibitemOpen
  \bibfield  {author} {\bibinfo {author} {\bibfnamefont {J.}~\bibnamefont {Kim}}, \bibinfo {author} {\bibfnamefont {J.}~\bibnamefont {Murugan}}, \bibinfo {author} {\bibfnamefont {J.}~\bibnamefont {Olle}},\ and\ \bibinfo {author} {\bibfnamefont {D.}~\bibnamefont {Rosa}},\ }\bibfield  {title} {\bibinfo {title} {Operator delocalization in quantum networks},\ }\href@noop {} {\bibfield  {journal} {\bibinfo  {journal} {Phys. Rev. A}\ }\textbf {\bibinfo {volume} {105}},\ \bibinfo {pages} {L010201} (\bibinfo {year} {2022})}\BibitemShut {NoStop}%
\bibitem [{\citenamefont {Cohen}(2017)}]{cohen2017thermalization}%
  \BibitemOpen
  \bibfield  {author} {\bibinfo {author} {\bibfnamefont {R.~Y.}\ \bibnamefont {Cohen}},\ }\emph {\bibinfo {title} {Thermalization of a 1-dimensional Rydberg gas and entanglement distribution across quantum networks}},\ \href@noop {} {Ph.D. thesis},\ \bibinfo  {school} {Universit{\'e} Paris Saclay (COmUE)} (\bibinfo {year} {2017})\BibitemShut {NoStop}%
\bibitem [{\citenamefont {Yu}\ and\ \citenamefont {Eberly}(2006)}]{yu2006quantum}%
  \BibitemOpen
  \bibfield  {author} {\bibinfo {author} {\bibfnamefont {T.}~\bibnamefont {Yu}}\ and\ \bibinfo {author} {\bibfnamefont {J.}~\bibnamefont {Eberly}},\ }\bibfield  {title} {\bibinfo {title} {Quantum open system theory: bipartite aspects},\ }\href@noop {} {\bibfield  {journal} {\bibinfo  {journal} {Phys. Rev. Lett.}\ }\textbf {\bibinfo {volume} {97}},\ \bibinfo {pages} {140403} (\bibinfo {year} {2006})}\BibitemShut {NoStop}%
\bibitem [{\citenamefont {Lankinen}\ \emph {et~al.}(2016{\natexlab{a}})\citenamefont {Lankinen}, \citenamefont {Lyyra}, \citenamefont {Sokolov}, \citenamefont {Teittinen}, \citenamefont {Ziaei},\ and\ \citenamefont {Maniscalco}}]{lankinen2016erratum}%
  \BibitemOpen
  \bibfield  {author} {\bibinfo {author} {\bibfnamefont {J.}~\bibnamefont {Lankinen}}, \bibinfo {author} {\bibfnamefont {H.}~\bibnamefont {Lyyra}}, \bibinfo {author} {\bibfnamefont {B.}~\bibnamefont {Sokolov}}, \bibinfo {author} {\bibfnamefont {J.}~\bibnamefont {Teittinen}}, \bibinfo {author} {\bibfnamefont {B.}~\bibnamefont {Ziaei}},\ and\ \bibinfo {author} {\bibfnamefont {S.}~\bibnamefont {Maniscalco}},\ }\bibfield  {title} {\bibinfo {title} {Erratum: Complete positivity, finite-temperature effects, and additivity of noise for time-local qubit dynamics [phys. rev. a 93, 052103 (2016)]},\ }\href@noop {} {\bibfield  {journal} {\bibinfo  {journal} {Phys. Rev. A}\ }\textbf {\bibinfo {volume} {94}},\ \bibinfo {pages} {059904} (\bibinfo {year} {2016}{\natexlab{a}})}\BibitemShut {NoStop}%
\bibitem [{\citenamefont {Jahromi}\ \emph {et~al.}(2020)\citenamefont {Jahromi}, \citenamefont {Mahdavipour}, \citenamefont {Khazaei~Shadfar},\ and\ \citenamefont {Lo~Franco}}]{jahromi2020witnessing}%
  \BibitemOpen
  \bibfield  {author} {\bibinfo {author} {\bibfnamefont {H.~R.}\ \bibnamefont {Jahromi}}, \bibinfo {author} {\bibfnamefont {K.}~\bibnamefont {Mahdavipour}}, \bibinfo {author} {\bibfnamefont {M.}~\bibnamefont {Khazaei~Shadfar}},\ and\ \bibinfo {author} {\bibfnamefont {R.}~\bibnamefont {Lo~Franco}},\ }\bibfield  {title} {\bibinfo {title} {Witnessing non-markovian effects of quantum processes through hilbert-schmidt speed},\ }\href@noop {} {\bibfield  {journal} {\bibinfo  {journal} {Phys. Rev. A}\ }\textbf {\bibinfo {volume} {102}},\ \bibinfo {pages} {022221} (\bibinfo {year} {2020})}\BibitemShut {NoStop}%
\bibitem [{\citenamefont {Mortezapour}\ and\ \citenamefont {Lo~Franco}(2018)}]{mortezapour2018protecting}%
  \BibitemOpen
  \bibfield  {author} {\bibinfo {author} {\bibfnamefont {A.}~\bibnamefont {Mortezapour}}\ and\ \bibinfo {author} {\bibfnamefont {R.}~\bibnamefont {Lo~Franco}},\ }\bibfield  {title} {\bibinfo {title} {Protecting quantum resources via frequency modulation of qubits in leaky cavities},\ }\href@noop {} {\bibfield  {journal} {\bibinfo  {journal} {Sci. rep.}\ }\textbf {\bibinfo {volume} {8}},\ \bibinfo {pages} {14304} (\bibinfo {year} {2018})}\BibitemShut {NoStop}%
\bibitem [{\citenamefont {Breuer}\ \emph {et~al.}(2016)\citenamefont {Breuer}, \citenamefont {Laine}, \citenamefont {Piilo},\ and\ \citenamefont {Vacchini}}]{breuer2016colloquium}%
  \BibitemOpen
  \bibfield  {author} {\bibinfo {author} {\bibfnamefont {H.-P.}\ \bibnamefont {Breuer}}, \bibinfo {author} {\bibfnamefont {E.-M.}\ \bibnamefont {Laine}}, \bibinfo {author} {\bibfnamefont {J.}~\bibnamefont {Piilo}},\ and\ \bibinfo {author} {\bibfnamefont {B.}~\bibnamefont {Vacchini}},\ }\bibfield  {title} {\bibinfo {title} {Colloquium: Non-markovian dynamics in open quantum systems},\ }\href@noop {} {\bibfield  {journal} {\bibinfo  {journal} {Rev. Mod. Phys.}\ }\textbf {\bibinfo {volume} {88}},\ \bibinfo {pages} {021002} (\bibinfo {year} {2016})}\BibitemShut {NoStop}%
\bibitem [{\citenamefont {Nielsen}\ and\ \citenamefont {Chuang}(2010)}]{nielsen2010quantum}%
  \BibitemOpen
  \bibfield  {author} {\bibinfo {author} {\bibfnamefont {M.~A.}\ \bibnamefont {Nielsen}}\ and\ \bibinfo {author} {\bibfnamefont {I.~L.}\ \bibnamefont {Chuang}},\ }\href@noop {} {\emph {\bibinfo {title} {Quantum computation and quantum information}}}\ (\bibinfo  {publisher} {Cambridge university press},\ \bibinfo {year} {2010})\BibitemShut {NoStop}%
\bibitem [{\citenamefont {He}\ \emph {et~al.}(2019)\citenamefont {He}, \citenamefont {Zeng}, \citenamefont {Chen},\ and\ \citenamefont {Yao}}]{he2019non}%
  \BibitemOpen
  \bibfield  {author} {\bibinfo {author} {\bibfnamefont {Z.}~\bibnamefont {He}}, \bibinfo {author} {\bibfnamefont {H.-S.}\ \bibnamefont {Zeng}}, \bibinfo {author} {\bibfnamefont {Y.}~\bibnamefont {Chen}},\ and\ \bibinfo {author} {\bibfnamefont {C.}~\bibnamefont {Yao}},\ }\bibfield  {title} {\bibinfo {title} {Non-markovian dynamics of a dephasing model in a squeezed thermal bath},\ }\href@noop {} {\bibfield  {journal} {\bibinfo  {journal} {Laser Phys. Lett.}\ }\textbf {\bibinfo {volume} {16}},\ \bibinfo {pages} {065204} (\bibinfo {year} {2019})}\BibitemShut {NoStop}%
\bibitem [{\citenamefont {Lankinen}\ \emph {et~al.}(2016{\natexlab{b}})\citenamefont {Lankinen}, \citenamefont {Lyyra}, \citenamefont {Sokolov}, \citenamefont {Teittinen}, \citenamefont {Ziaei},\ and\ \citenamefont {Maniscalco}}]{lankinen2016complete}%
  \BibitemOpen
  \bibfield  {author} {\bibinfo {author} {\bibfnamefont {J.}~\bibnamefont {Lankinen}}, \bibinfo {author} {\bibfnamefont {H.}~\bibnamefont {Lyyra}}, \bibinfo {author} {\bibfnamefont {B.}~\bibnamefont {Sokolov}}, \bibinfo {author} {\bibfnamefont {J.}~\bibnamefont {Teittinen}}, \bibinfo {author} {\bibfnamefont {B.}~\bibnamefont {Ziaei}},\ and\ \bibinfo {author} {\bibfnamefont {S.}~\bibnamefont {Maniscalco}},\ }\bibfield  {title} {\bibinfo {title} {Complete positivity, finite-temperature effects, and additivity of noise for time-local qubit dynamics},\ }\href@noop {} {\bibfield  {journal} {\bibinfo  {journal} {Phys. Rev. A}\ }\textbf {\bibinfo {volume} {93}},\ \bibinfo {pages} {052103} (\bibinfo {year} {2016}{\natexlab{b}})}\BibitemShut {NoStop}%
\bibitem [{\citenamefont {Scully}\ and\ \citenamefont {Zubairy}(1999)}]{scully1999quantum}%
  \BibitemOpen
  \bibfield  {author} {\bibinfo {author} {\bibfnamefont {M.~O.}\ \bibnamefont {Scully}}\ and\ \bibinfo {author} {\bibfnamefont {M.~S.}\ \bibnamefont {Zubairy}},\ }\href@noop {} {\bibinfo {title} {Quantum optics}} (\bibinfo {year} {1999})\BibitemShut {NoStop}%
\bibitem [{\citenamefont {Reina}\ \emph {et~al.}(2002)\citenamefont {Reina}, \citenamefont {Quiroga},\ and\ \citenamefont {Johnson}}]{reina2002decoherence}%
  \BibitemOpen
  \bibfield  {author} {\bibinfo {author} {\bibfnamefont {J.~H.}\ \bibnamefont {Reina}}, \bibinfo {author} {\bibfnamefont {L.}~\bibnamefont {Quiroga}},\ and\ \bibinfo {author} {\bibfnamefont {N.~F.}\ \bibnamefont {Johnson}},\ }\bibfield  {title} {\bibinfo {title} {Decoherence of quantum registers},\ }\href@noop {} {\bibfield  {journal} {\bibinfo  {journal} {Phys. Rev. A}\ }\textbf {\bibinfo {volume} {65}},\ \bibinfo {pages} {032326} (\bibinfo {year} {2002})}\BibitemShut {NoStop}%
\bibitem [{\citenamefont {Chanda}\ and\ \citenamefont {Bhattacharya}(2016)}]{chanda2016delineating}%
  \BibitemOpen
  \bibfield  {author} {\bibinfo {author} {\bibfnamefont {T.}~\bibnamefont {Chanda}}\ and\ \bibinfo {author} {\bibfnamefont {S.}~\bibnamefont {Bhattacharya}},\ }\bibfield  {title} {\bibinfo {title} {Delineating incoherent non-markovian dynamics using quantum coherence},\ }\href@noop {} {\bibfield  {journal} {\bibinfo  {journal} {Annals of Physics}\ }\textbf {\bibinfo {volume} {366}},\ \bibinfo {pages} {1} (\bibinfo {year} {2016})}\BibitemShut {NoStop}%
\bibitem [{\citenamefont {Teittinen}\ \emph {et~al.}(2018)\citenamefont {Teittinen}, \citenamefont {Lyyra}, \citenamefont {Sokolov},\ and\ \citenamefont {Maniscalco}}]{teittinen2018revealing}%
  \BibitemOpen
  \bibfield  {author} {\bibinfo {author} {\bibfnamefont {J.}~\bibnamefont {Teittinen}}, \bibinfo {author} {\bibfnamefont {H.}~\bibnamefont {Lyyra}}, \bibinfo {author} {\bibfnamefont {B.}~\bibnamefont {Sokolov}},\ and\ \bibinfo {author} {\bibfnamefont {S.}~\bibnamefont {Maniscalco}},\ }\bibfield  {title} {\bibinfo {title} {Revealing memory effects in phase-covariant quantum master equations},\ }\href@noop {} {\bibfield  {journal} {\bibinfo  {journal} {New Journal of Physics}\ }\textbf {\bibinfo {volume} {20}},\ \bibinfo {pages} {073012} (\bibinfo {year} {2018})}\BibitemShut {NoStop}%
\bibitem [{\citenamefont {Palma}\ \emph {et~al.}(1996)\citenamefont {Palma}, \citenamefont {Suominen},\ and\ \citenamefont {Ekert}}]{palma1996quantum}%
  \BibitemOpen
  \bibfield  {author} {\bibinfo {author} {\bibfnamefont {G.~M.}\ \bibnamefont {Palma}}, \bibinfo {author} {\bibfnamefont {K.-A.}\ \bibnamefont {Suominen}},\ and\ \bibinfo {author} {\bibfnamefont {A.}~\bibnamefont {Ekert}},\ }\bibfield  {title} {\bibinfo {title} {Quantum computers and dissipation},\ }\href@noop {} {\bibfield  {journal} {\bibinfo  {journal} {Proc. R. Soc. London A}\ }\textbf {\bibinfo {volume} {452}},\ \bibinfo {pages} {567} (\bibinfo {year} {1996})}\BibitemShut {NoStop}%
\bibitem [{\citenamefont {Haikka}\ \emph {et~al.}(2013)\citenamefont {Haikka}, \citenamefont {Johnson},\ and\ \citenamefont {Maniscalco}}]{haikka2013non}%
  \BibitemOpen
  \bibfield  {author} {\bibinfo {author} {\bibfnamefont {P.}~\bibnamefont {Haikka}}, \bibinfo {author} {\bibfnamefont {T.}~\bibnamefont {Johnson}},\ and\ \bibinfo {author} {\bibfnamefont {S.}~\bibnamefont {Maniscalco}},\ }\bibfield  {title} {\bibinfo {title} {Non-markovianity of local dephasing channels and time-invariant discord},\ }\href@noop {} {\bibfield  {journal} {\bibinfo  {journal} {Phys. Rev. A}\ }\textbf {\bibinfo {volume} {87}},\ \bibinfo {pages} {010103} (\bibinfo {year} {2013})}\BibitemShut {NoStop}%
\bibitem [{\citenamefont {Tuorila}\ \emph {et~al.}(2010)\citenamefont {Tuorila}, \citenamefont {Silveri}, \citenamefont {Sillanp{\"a}{\"a}}, \citenamefont {Thuneberg}, \citenamefont {Makhlin},\ and\ \citenamefont {Hakonen}}]{tuorila2010stark}%
  \BibitemOpen
  \bibfield  {author} {\bibinfo {author} {\bibfnamefont {J.}~\bibnamefont {Tuorila}}, \bibinfo {author} {\bibfnamefont {M.}~\bibnamefont {Silveri}}, \bibinfo {author} {\bibfnamefont {M.}~\bibnamefont {Sillanp{\"a}{\"a}}}, \bibinfo {author} {\bibfnamefont {E.}~\bibnamefont {Thuneberg}}, \bibinfo {author} {\bibfnamefont {Y.}~\bibnamefont {Makhlin}},\ and\ \bibinfo {author} {\bibfnamefont {P.}~\bibnamefont {Hakonen}},\ }\bibfield  {title} {\bibinfo {title} {Stark effect and generalized bloch-siegert shift in a strongly driven two-level system},\ }\href@noop {} {\bibfield  {journal} {\bibinfo  {journal} {Phys. Rev. Lett.}\ }\textbf {\bibinfo {volume} {105}},\ \bibinfo {pages} {257003} (\bibinfo {year} {2010})}\BibitemShut {NoStop}%
\bibitem [{\citenamefont {Nakamura}\ \emph {et~al.}(2001)\citenamefont {Nakamura}, \citenamefont {Pashkin},\ and\ \citenamefont {Tsai}}]{nakamura2001rabi}%
  \BibitemOpen
  \bibfield  {author} {\bibinfo {author} {\bibfnamefont {Y.}~\bibnamefont {Nakamura}}, \bibinfo {author} {\bibfnamefont {Y.~A.}\ \bibnamefont {Pashkin}},\ and\ \bibinfo {author} {\bibfnamefont {J.~S.}\ \bibnamefont {Tsai}},\ }\bibfield  {title} {\bibinfo {title} {Rabi oscillations in a josephson-junction charge two-level system},\ }\href@noop {} {\bibfield  {journal} {\bibinfo  {journal} {Phys. Rev. Lett.}\ }\textbf {\bibinfo {volume} {87}},\ \bibinfo {pages} {246601} (\bibinfo {year} {2001})}\BibitemShut {NoStop}%
\bibitem [{\citenamefont {Li}\ \emph {et~al.}(2013)\citenamefont {Li}, \citenamefont {Silveri}, \citenamefont {Kumar}, \citenamefont {Pirkkalainen}, \citenamefont {Veps{\"a}l{\"a}inen}, \citenamefont {Chien}, \citenamefont {Tuorila}, \citenamefont {Sillanp{\"a}{\"a}}, \citenamefont {Hakonen}, \citenamefont {Thuneberg} \emph {et~al.}}]{li2013motional}%
  \BibitemOpen
  \bibfield  {author} {\bibinfo {author} {\bibfnamefont {J.}~\bibnamefont {Li}}, \bibinfo {author} {\bibfnamefont {M.}~\bibnamefont {Silveri}}, \bibinfo {author} {\bibfnamefont {K.}~\bibnamefont {Kumar}}, \bibinfo {author} {\bibfnamefont {J.-M.}\ \bibnamefont {Pirkkalainen}}, \bibinfo {author} {\bibfnamefont {A.}~\bibnamefont {Veps{\"a}l{\"a}inen}}, \bibinfo {author} {\bibfnamefont {W.}~\bibnamefont {Chien}}, \bibinfo {author} {\bibfnamefont {J.}~\bibnamefont {Tuorila}}, \bibinfo {author} {\bibfnamefont {M.}~\bibnamefont {Sillanp{\"a}{\"a}}}, \bibinfo {author} {\bibfnamefont {P.}~\bibnamefont {Hakonen}}, \bibinfo {author} {\bibfnamefont {E.}~\bibnamefont {Thuneberg}}, \emph {et~al.},\ }\bibfield  {title} {\bibinfo {title} {Motional averaging in a superconducting qubit},\ }\href@noop {} {\bibfield  {journal} {\bibinfo  {journal} {Nat. Commun.}\ }\textbf {\bibinfo {volume} {4}},\ \bibinfo {pages} {1420} (\bibinfo {year} {2013})}\BibitemShut {NoStop}%
\bibitem [{\citenamefont {Oliver}\ \emph {et~al.}(2005)\citenamefont {Oliver}, \citenamefont {Yu}, \citenamefont {Lee}, \citenamefont {Berggren}, \citenamefont {Levitov},\ and\ \citenamefont {Orlando}}]{oliver2005mach}%
  \BibitemOpen
  \bibfield  {author} {\bibinfo {author} {\bibfnamefont {W.~D.}\ \bibnamefont {Oliver}}, \bibinfo {author} {\bibfnamefont {Y.}~\bibnamefont {Yu}}, \bibinfo {author} {\bibfnamefont {J.~C.}\ \bibnamefont {Lee}}, \bibinfo {author} {\bibfnamefont {K.~K.}\ \bibnamefont {Berggren}}, \bibinfo {author} {\bibfnamefont {L.~S.}\ \bibnamefont {Levitov}},\ and\ \bibinfo {author} {\bibfnamefont {T.~P.}\ \bibnamefont {Orlando}},\ }\bibfield  {title} {\bibinfo {title} {Mach-zehnder interferometry in a strongly driven superconducting qubit},\ }\href@noop {} {\bibfield  {journal} {\bibinfo  {journal} {Science}\ }\textbf {\bibinfo {volume} {310}},\ \bibinfo {pages} {1653} (\bibinfo {year} {2005})}\BibitemShut {NoStop}%
\bibitem [{\citenamefont {Cao}\ \emph {et~al.}(2013)\citenamefont {Cao}, \citenamefont {Li}, \citenamefont {Tu}, \citenamefont {Wang}, \citenamefont {Zhou}, \citenamefont {Xiao}, \citenamefont {Guo}, \citenamefont {Jiang},\ and\ \citenamefont {Guo}}]{cao2013ultrafast}%
  \BibitemOpen
  \bibfield  {author} {\bibinfo {author} {\bibfnamefont {G.}~\bibnamefont {Cao}}, \bibinfo {author} {\bibfnamefont {H.-O.}\ \bibnamefont {Li}}, \bibinfo {author} {\bibfnamefont {T.}~\bibnamefont {Tu}}, \bibinfo {author} {\bibfnamefont {L.}~\bibnamefont {Wang}}, \bibinfo {author} {\bibfnamefont {C.}~\bibnamefont {Zhou}}, \bibinfo {author} {\bibfnamefont {M.}~\bibnamefont {Xiao}}, \bibinfo {author} {\bibfnamefont {G.-C.}\ \bibnamefont {Guo}}, \bibinfo {author} {\bibfnamefont {H.-W.}\ \bibnamefont {Jiang}},\ and\ \bibinfo {author} {\bibfnamefont {G.-P.}\ \bibnamefont {Guo}},\ }\bibfield  {title} {\bibinfo {title} {Ultrafast universal quantum control of a quantum-dot charge qubit using landau--zener--st{\"u}ckelberg interference},\ }\href@noop {} {\bibfield  {journal} {\bibinfo  {journal} {Nat. Commun.}\ }\textbf {\bibinfo {volume} {4}},\ \bibinfo {pages} {1401} (\bibinfo {year} {2013})}\BibitemShut {NoStop}%
\bibitem [{\citenamefont {Trabesinger}(2017)}]{trabesinger2017quantum}%
  \BibitemOpen
  \bibfield  {author} {\bibinfo {author} {\bibfnamefont {A.}~\bibnamefont {Trabesinger}},\ }\bibfield  {title} {\bibinfo {title} {Quantum computing: towards reality},\ }\href@noop {} {\bibfield  {journal} {\bibinfo  {journal} {Nature}\ }\textbf {\bibinfo {volume} {543}},\ \bibinfo {pages} {S1} (\bibinfo {year} {2017})}\BibitemShut {NoStop}%
\end{thebibliography}%

\end{document}